\documentclass[twocolumn,journal]{IEEEtran}
\pdfoutput=1
\usepackage[T1]{fontenc}
\usepackage{xcolor}
\usepackage{pdfcolmk}
\usepackage{float}
\usepackage{amsmath}
\usepackage{amssymb}
\usepackage{graphicx}
\usepackage{setspace}

\makeatletter

\providecommand{\tabularnewline}{\\}
\floatstyle{ruled}
\newfloat{algorithm}{tbp}{loa}
\providecommand{\algorithmname}{Algorithm}
\floatname{algorithm}{\protect\algorithmname}
\providecolor{lyxadded}{rgb}{0,0,1}
\providecolor{lyxdeleted}{rgb}{1,0,0}
\DeclareRobustCommand{\lyxadded}[3]{{\texorpdfstring{\color{lyxadded}{}}{}#3}}

 \usepackage{changebar}
 \providecommand{\lyxadded}[3]{}
 
 \renewcommand{\lyxadded}[3]{
   {\protect\cbstart\color{lyxadded}{}#3\protect\cbend}
 }

\usepackage[caption=false,font=footnotesize]{subfig}
 \usepackage{cite}

\makeatother
\begin{document}

\title{Performance Analysis and Compensation of Joint TX/RX I/Q Imbalance
in Differential STBC-OFDM}

\author{\IEEEauthorblockN{Lei Chen, Ahmed G. Helmy, Guangrong Yue, Shaoqian
Li, \textit{Fellow, IEEE,} and Naofal Al-Dhahir, \textit{Fellow, IEEE}{\small{}}
}\\
\thanks{Copyright (c) 2015 IEEE. Personal use of this material is permitted.
However, permission to use this material for any other purposes must
be obtained from the IEEE by sending a request to pub-spermissions@ieee.org.
\protect \par
L. Chen, G. Yue and S. Li are with the National Key Laboratory of Science and Technology on Communications,
University of Electronic Science and Technology of China, Chengdu 611731,
China: leichen1008@hotmail.com, \{yuegr,lsq\}@uestc.edu.cn. \protect \par
A. G. Helmy and N. Al-Dhahir are with The University of Texas at Dallas,
TX, USA, \{ahmed.g.helmy, aldhahir\}@utdallas.edu.\protect \par
This work was done while Lei Chen was a visiting PhD student at University
of Texas at Dallas and was supported in part by the scholarship from
China Scholarship Council (CSC). The work of A. Helmy and N. Al-Dhahir was made possible by NPRP grant \#NPRP 8-627-2-260 from the Qatar
National Research Fund (a member of Qatar Foundation). The work of L. Chen, G. Yue and S. Li is supported by the National Science and Technology Major Project (No.2013ZX03005010), National Natural Science Foundation of China (No.61371103, No.61401447), ITDCN open program (No.KX152600016/ITD-U15007). The statements made herein are solely the responsibility of the authors. }}
\maketitle
\begin{abstract}
Differential space time block coding (STBC) achieves full spatial
diversity and avoids channel estimation overhead. Over highly frequency-selective
channels, STBC is integrated with orthogonal frequency division multiplexing
(OFDM) to efficiently mitigate intersymbol interference effects. However,
low-cost implementation of STBC-OFDM with direct-conversion transceivers
is sensitive to In-phase/Quadrature-phase imbalance (IQI). In this
paper, we quantify the performance impact of IQI at both the transmitter
and receiver radio frequency front-ends on differential STBC-OFDM
systems which has not been investigated before in the literature.
In addition, we propose a widely-linear compensation algorithm at
the receiver to mitigate the performance degradation caused by the
IQI at the transmitter and receiver ends. Moreover, a parameter-based
generalized algorithm is proposed to extract the IQI parameters and
improve the performance under high-mobility. The adaptive compensation
algorithms are blind and work in a decision-directed manner without
using known pilots or training sequences. Numerical results show that
our proposed compensation algorithms can effectively mitigate IQI
in differential STBC-OFDM.
\end{abstract}


\IEEEpeerreviewmaketitle{}

\section{Introduction}

\vspace{-.2cm}\IEEEPARstart{S}{pace-time} block-coded orthogonal
frequency-division multiplexing (STBC-OFDM) is an effective transceiver
structure to mitigate the wireless channel's frequency selectivity
while realizing multipaths and spatial diversity gains \cite{ProceedingNaofal}.
To acquire channel knowledge for signal detection at the receiver,
STBC-OFDM schemes require the transmission of pilot signals \cite{Tarokh2000}.
However, to avoid the rate loss due to pilot signal overhead, we may
want to forego channel estimation to avoid its complexity and the
degradation of channel tracking quality in a fast time-varying environment
\cite{Diggavi2002,S_luDiff}. Differential transmission/detection,
which is adopted in several standards such as digital audio broadcasting
(DAB) \cite{etsi1995300}, achieves this goal and has been successfully
integrated with STBC-OFDM \cite{Tarokh2000,Diggavi2002,Hughes2000}.\textcolor{red}{{}
}Although a differential STBC-OFDM system avoids the overhead of channel
estimation, a low-cost transceiver implementation based on the direct-conversion
architecture suffers from analog and radio-frequency (RF) impairments.
The impairments in the analog components are mainly due to the uncontrollable
fabrication process variations. Since most of these impairments cannot
be effectively eliminated in the analog domain, an efficient compensation
algorithm in the digital baseband domain would be highly desirable.
One of the main sources of the analog impairments is the imbalance
between the In-phase (I) and Quadrature-phase (Q) branches when the
transmitted signal is up-converted to RF frequency at the transmitter,
or the received RF signal is down-converted to baseband at the receiver.
The I/Q imbalance (IQI) arises due to mismatches between the I and
Q branches from the ideal case, $i.e.$, from the exact $90^{o}$
phase difference and equal amplitudes between the sine and cosine
branches. In OFDM systems, IQI destroys the subcarrier orthogonality
by introducing inter-carrier interference (ICI) between mirror subcarriers
which can lead to serious performance degradation \cite{Tarighat2005}. 

Many papers investigated IQI in single-input single-output OFDM systems
(see \cite{Tarighat2005,Tarighat2007,7286857,7265096}
and the references therein). There are also several works dealing
with IQI in coherent multiple-antenna systems. In \cite{narasimhan2010reduced},
a super-block structure for the Alamouti STBC scheme is designed to
ensure orthogonality in the presence of IQI. In \cite{Jointestimationtransmitter_Marey_2013},
an Expectation-Maximization-based algorithm is proposed to mitigate
IQI in Alamouti-based STBC-OFDM systems. An equalization algorithm
is proposed to mitigate IQI in STBC-OFDM systems in \cite{STBCMIMOOFDM_Tandur_2008}.
The authors in \cite{PerformanceAnalysisSpace_Zou_2007} analyze and
compensate for IQI in single-carrier STBC systems. Precoding methods
were investigated in coherent massive MIMO systems in \cite{7063451}.
In \cite{7006774}, I/Q imbalance effects were left uncompensated;
instead, a link adaption strategy was considered and an IQI-aware
transmission method was developed. In \cite{7458826}, an IQI-robust
channel estimation and low-complexity compensation method for RX-IQI
was proposed.

Although the compensation of IQI in STBC-OFDM systems has been well
investigated, all of the existing works deal with IQI in a coherent
system, where the channel state information (CSI) is assumed known
or estimated at the receiver. \textbf{To the best of our knowledge,
there is no previous work dealing with joint transmitter and receiver
IQI in differential transmission systems}. Note that some blind or
semi-blind estimation schemes that do not require CSI for IQI compensation
have been proposed in the literature \cite{GaoZhuLinEtAl2010,ZouValkamaRenfors2008,WittRooyen2009,RykaczewskiJondral2007}.
A semi-blind compensation algorithm for IQI in MIMO-OFDM systems is
presented in \cite{GaoZhuLinEtAl2010}, where a blind signal separation
(BSS) method is used to equalize the equivalent channel including
IQI and the wireless multipath channel. However a known reference
signal must be embedded in the transmitted signal. Another BSS-based
IQI compensation method is presented in \cite{ZouValkamaRenfors2008},
and blind estimation methods of transmitter IQI (TX-IQI) and receiver
IQI (RX-IQI) are presented in \cite{WittRooyen2009}, \cite{RykaczewskiJondral2007}.
The difference between the compensation algorithm in this paper and
the blind compensation algorithms in \cite{GaoZhuLinEtAl2010,ZouValkamaRenfors2008,WittRooyen2009,RykaczewskiJondral2007}
is that blind compensation algorithms do not make use of the differential
encoding property and they typically suffer from local optima and
very slow convergence. On the other hand, although some of these blind
compensation algorithms estimate IQI in a blind manner \cite{ZouValkamaRenfors2008,WittRooyen2009,RykaczewskiJondral2007},
they require equalization before the compensation of transmitter IQI,
which is not feasible in a system with differential detection. 

We extend the work in \cite{gc} which only analyzes and compensates
for the effect of the RX-IQI in the differential STBC-OFDM (DSTBC-OFDM)
transmission. In this paper, we analyze the joint
impact of TX-IQI and RX-IQI in DSTBC-OFDM systems. An equivalent
signal power degradation factor due to TX-IQI is derived. Moreover,
by using the differential encoding property of the transmitted signal,
we propose an adaptive decision-directed algorithm that uses a widely-linear
(WL) structure to compensate for TX-IQI and RX-IQI without the need
for knowing or estimating the CSI. Additionally, we propose a parameter-based
(PB) generalized algorithm that can enhance the compensation performance
under high-mobility. The rest of this paper is organized as follows:
the system model of DSTBC-OFDM in the presence of TX-IQI and RX-IQI
is developed in Section \ref{sec:System-Model}. In Section \ref{sec:ber},
we discuss the impact of IQI on the bit error rate (BER) performance
of DSTBC-OFDM. We propose a decision-directed IQI compensation algorithm
in Section \ref{sec:compensateion} and the numerical results are
presented in Section \ref{sec:numerical}. Finally, we conclude our
paper in Section \ref{sec:conc}. 

\textit{Notations}: Unless further noted, frequency-domain (FD) matrices
and vectors are denoted by upper-case and lower-case boldface, respectively.
Time-domain (TD) matrices and vectors are denoted by upper-case and
lower-case boldface with under-bar, respectively. We denote the Hermitian,
$i.e.$ complex-conjugate transpose of a matrix or a vector by ${{(\cdot)}^{H}}$.
The conjugate and transpose of a matrix, a vector, or a scalar is
denoted by ${{(\cdot)}^{*}}$ and ${{(\cdot)}^{T}}$, respectively.
The symbol ${{[\mathbf{A}]}_{m,n}}$ denotes the entry at the $m$-th
row and the $n$-th column of matrix $\mathbf{A}$. Matrix $\mathbf{F}$
is the $N$-point Discrete Fourier Transform (DFT) matrix whose entries
are given by: ${{[\mathbf{F}]}_{m,n}}=\frac{1}{\sqrt{N}}\exp(-j\frac{2\pi}{N}mn)$,
with $0\le m,n\le N-1$. $\textrm{Re}\{\cdot\}$ and $\textrm{Im}\{\cdot\}$
denote the real and imaginary parts of a complex number, respectively.
For convenience, Table \ref{tab:List-of-Key-1} summarizes the key
variables used throughout the paper.\vspace{0.3cm}

\begin{table}[tb]
\caption{List of Key Variables\label{tab:List-of-Key-1}}

{\small{}}%
\begin{tabular}{|c|c|}
\hline 
{\small{}Var.} & {\small{}Definition}\tabularnewline
\hline 
\hline 
{\small{}$N$} & {\small{}Number of subcarriers} \tabularnewline
\hline  {\small{}$\rho_{t/r}$} & {\small{}Interference-to-signal ratio of TX/RX-IQI}\tabularnewline
\hline 
{\small{}$M$} & {\small{}Modulation order of PSK signal}\tabularnewline
\hline  {\small{}$\mathbf{\Gamma}_{c}(n)$} & {\small{}Compensation matrix for the $n$-th subcarrier}\tabularnewline
\hline 
{\small{}$\mathbf{U}_{k}(n)$} & {\small{}$k$-th FD information block on the $n$-th subcarrier} \tabularnewline
\hline  {\small{}$\lambda_{i}(n)$} & {\small{}FD channel of Subcarrier $n$ from the $i$-th transmit antenna}\tabularnewline
\hline 
{\small{}$\mathbf{S}_{k}\left(n\right)$} & {\small{}$k$-th FD STBC block on the $n$-th subcarrier} \tabularnewline
\hline  {\small{}$g_{t/r}$} & {\small{}Amplitude imbalance of TX/RX}\tabularnewline
\hline 
{\small{}$\mathbf{\bar{S}}_{k}$} & {\small{}$k$-th FD STBC block on the $(N-n+2)$-th subcarrier} \tabularnewline
\hline  {\small{}$\phi_{t/r}$ } & {\small{}Phase imbalance of TX/RX}\tabularnewline
\hline 
\end{tabular}{\small \par}
\end{table}

\section{System Model\label{sec:System-Model}}

\begin{figure}[tb]
\centering\includegraphics[width=3.5in]{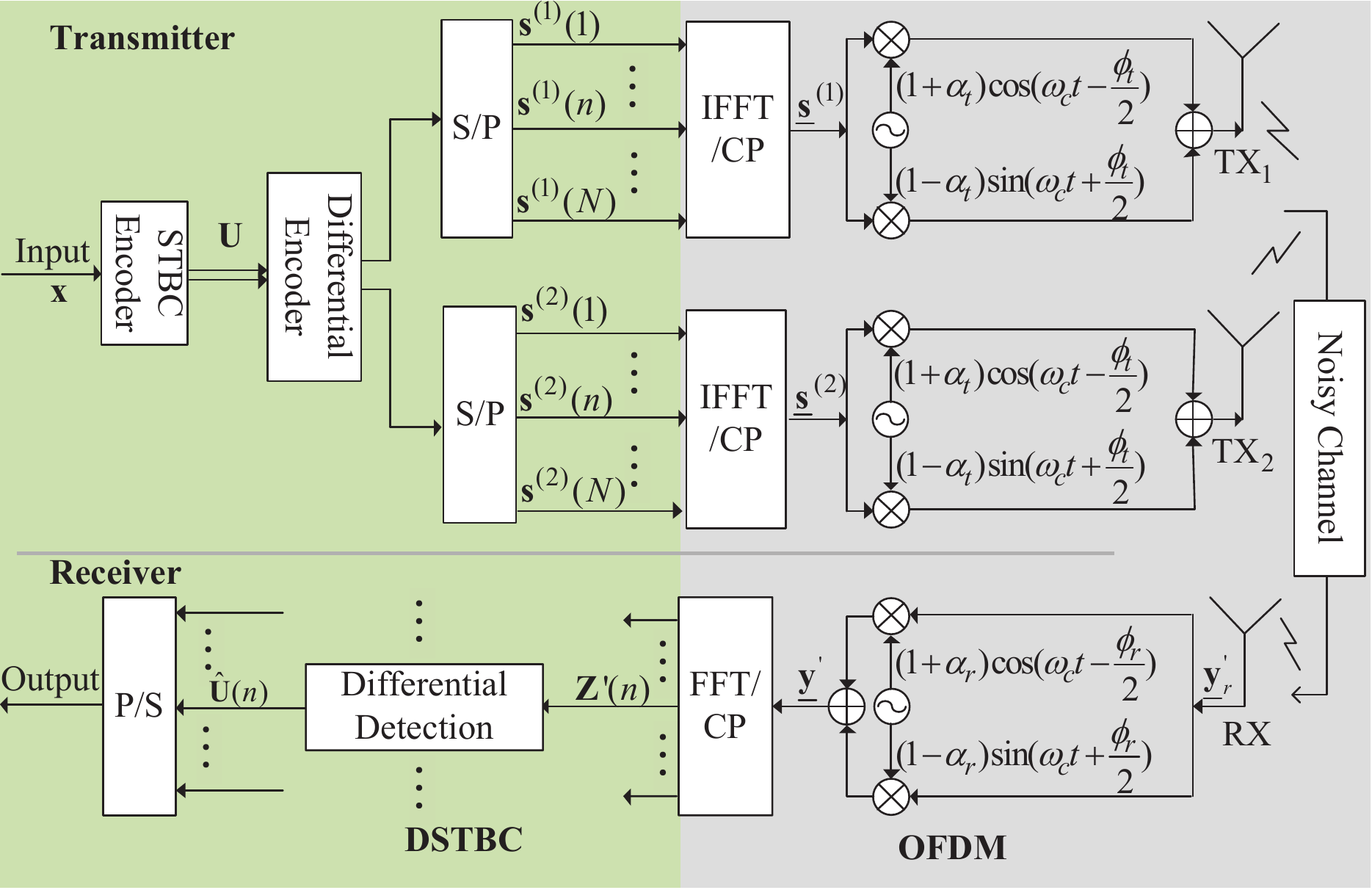} \caption{System model of DSTBC-OFDM under TX RX IQI\label{fig: sysmodel-1}}
\vspace{-0.5cm}
\end{figure}

In this section, we briefly introduce the system model of DSTBC-OFDM
under IQI depicted in Fig. \ref{fig: sysmodel-1}. Without loss of
generality, we consider an Alamouti-based \cite{Alamouti1998} differential
STBC-OFDM wireless communication system equipped with two transmit
antennas and a single receive antenna. For a multiple-antenna receiver,
the analysis and compensation algorithms in this paper can also be
straightforwardly applied. Since OFDM transmission divides the channel
into $N$ mutually orthogonal subcarriers, the OFDM-STBC input-output
model at the $n$-th subcarrier can be expressed as follows \cite{Diggavi2002}

{\small{}
\begin{eqnarray}
\mathbf{z}_{k}(n) & = & \left[\begin{matrix}\lambda_{1}(n) & \lambda_{2}(n)\end{matrix}\right]\underset{\mathbf{S}_{k}\left(n\right)}{\underbrace{\left[\begin{array}{cc}
s_{1}(k,n) & s_{2}(k,n)\\
-s_{2}^{*}(k,n) & s_{1}^{*}(k,n)
\end{array}\right]}}\nonumber \\
 &  & +\left[\begin{array}{cc}
v_{2k+1}(n) & v_{2k+2}(n)\end{array}\right],\label{eq:systemodelorg}
\end{eqnarray}
}where $\mathbf{z}_{k}(n)=\left[\begin{array}{cc}
z_{2k+1}(n) & z_{2k+2}(n)\end{array}\right]$ is the received vector corresponding to the $(2k+1)$-th and $(2k+2)$-th
received OFDM symbols, respectively. Moreover, the matrix $\mathbf{S}_{k}\left(n\right)$
is the $k$-th frequency-domain differentially-encoded STBC transmission
matrix at the $n$-th subcarrier, whose first and second columns are
transmitted over the $(2k-1)$-th and $2k$-th OFDM symbols, respectively.
The first and second rows of $\mathbf{S}_{k}\left(n\right)$ are transmitted
by the first and second transmit antennas, respectively. In addition,
$v_{2k+1}(n)$ and $v_{2k+1}(n)$ are the frequency-domain additive
noise symbols at the $n$-th subcarrier. The scalars $\lambda_{i}(n)$
($i=1,2$) correspond to the channel coefficients between the $i$-th
transmit antenna and the single receive antenna at the $n$-th subcarrier
where the vector $\boldsymbol{\lambda}_{i}=[\lambda_{i}(1),\lambda_{i}(2),...,\lambda_{i}(N)]$
is given by \cite{Tarighat2005}

\begin{equation}
\boldsymbol{\lambda}_{i}=\sqrt{N}\mathbf{F}^{H}\left[\begin{matrix}\mbox{ }\mathbf{\underline{h}}_{i}\\
\mathbf{0}_{\left(N-\left(L+1\right)\right)\times1}
\end{matrix}\right],\label{eq:lmda}
\end{equation}
and the vector $\mbox{ }\mathbf{\underline{h}}_{i}=\left[\underline{h}_{0}^{\left(i\right)}\,\underline{h}_{1}^{\left(i\right)}\,\cdots\,\underline{h}_{L}^{\left(i\right)}\right]^{T}$
is the time-domain channel impulse response (CIR) vector between the
$i$-th transmit antenna and the single receive antenna with $L+1$
independent taps.\par
The differential encoding rule used to generate $\mathbf{S}_{k}\left(n\right)$
in Eq. \eqref{eq:systemodelorg} is given by \cite{Diggavi2002} 
\begin{equation}
\mathbf{S}_{k+1}(n)=\mathbf{S}_{k}(n)\mathbf{U}_{k+1}(n),\label{difsu-1}
\end{equation}
where $\mathbf{U}_{k+1}(n)$ is a $2\times2$ Alamouti-structured
information matrix given by
\begin{equation}
\mathbf{U}_{k+1}(n)=\left[\begin{array}{cc}
u_{1}(k+1,n) & u_{2}(k+1,n)\\
-u_{2}^{*}(k+1,n) & u_{1}^{*}(k+1,n)
\end{array}\right],
\end{equation}
where the symbols $u_{1}(k+1,n)$ and $u_{2}(k+1,n)$ are drawn from
a constant-modulus signal constellation $\varPsi$ (usually M-ary
phase shift keying (M-PSK) symbols), which is required by a conventional
differential transceiver. Note that few researchers have investigated
using a non-constant-modulus signal constellation in a differential
transceiver\cite{hwang2003differential}\cite{xu2016soft}, where
further operations are needed to extract the amplitude information
at the receiver which is beyond the scope of this paper. Using the
same approach followed in \cite{Diggavi2002}, Eq. \eqref{eq:systemodelorg}
can be expressed in the form of $\mathbf{Z}_{k}\left(n\right)=\mathbf{\Lambda}\left(n\right)\mathbf{S}_{k}\left(n\right)+\mathbf{V}_{k}\left(n\right)$
as follows
\begin{equation}
{\small \begin{array}{c}
{{\bf{Z}}_k}\left( n \right) = \underbrace {\left[ {\begin{array}{*{20}{c}}
		{{\lambda _1}(n)}&{{\lambda _2}(n)}\\
		{ - \lambda _2^*(n)}&{\lambda _1^*(n)}
		\end{array}} \right]}_{{\bf{\Lambda }}\left( n \right)}\underbrace {\left[ {\begin{array}{*{20}{c}}
		{{s_1}(k,n)}&{{s_2}(k,n)}\\
		{ - s_2^*(k,n)}&{s_1^*(k,n)}
		\end{array}} \right]}_{{{\bf{S}}_k}\left( n \right)}\\
+ \underbrace {\left[ {\begin{array}{*{20}{c}}
		{{v_{2k + 1}}(n)}&{{v_{2k + 2}}(n)}\\
		{ - v_{2k + 2}^*(n)}&{v_{2k + 1}^*(n)}
		\end{array}} \right]}_{{{\bf{V}}_k}\left( n \right)}
\end{array}}
\label{eq:model2}
\end{equation}
{\small}
where $\bf{Z}_k$ is also an Alamouti matrix given by
\begin{equation}
{{\bf{Z}}_k}\left( n \right) = \left[ {\begin{array}{*{20}{c}}
	{{z_{2k + 1}}(n)}&{{z_{2k + 2}}(n)}\\
	{ - z_{2k + 2}^*(n)}&{z_{2k + 1}^*(n)}
	\end{array}} \right].
\end{equation}
Based on the STBC-OFDM model in Eq. \eqref{eq:model2} and the differential
encoding in Eq. (\ref{difsu-1}), the received signal blocks $\mathbf{Z}_{k}\left(n\right)$
and $\mathbf{Z}_{k+1}\left(n\right)$ are defined by
\begin{align}
\mathbf{Z}_{k}\left(n\right) & =\mathbf{\Lambda}(n)\mathbf{S}_{k}\left(n\right)+\mathbf{V}_{k}\left(n\right),\nonumber \\
\mathbf{Z}_{k+1}\left(n\right) & =\mathbf{\Lambda}(n)\mathbf{S}_{k+1}\left(n\right)+\mathbf{V}_{k+1}\left(n\right)\nonumber \\
 & =\mathbf{\Lambda}(n)\mathbf{S}_{k}(n)\mathbf{U}_{k+1}(n)+\mathbf{V}_{k+1}\left(n\right).\label{eq:freqDiff}
\end{align}
Therefore, the maximum likelihood (ML) decoder for the information
matrix $\mathbf{U}_{k+1}(n)$ is given by \cite{Hughes2000} 
\begin{equation}
\mathbf{\hat{U}}_{k+1}(n)=\underset{\mathbf{\quad\quad\quad\mathbf{U}}}{\mathop{\arg}\quad\mathop{\max}}\,\left\{ \mathbf{\mathbf{U}}^{H}\mathbf{Z}_{k}^{H}\left(n\right)\mathbf{Z}_{k+1}\left(n\right)\right\} ,\label{ML}
\end{equation}
where $\mathbf{\mathbf{U}}$ is chosen from the Alamouti matrix sets
formed by all possible information matrices.\par
Starting from the system model in Eqs. (\ref{eq:model2}) and (\ref{eq:freqDiff}),
we adopt the time-domain TX-IQI and RX-IQI model defined in \cite{Tarighat2005}.
For a given generalized time-domain baseband signal $\underline{b}\left(t\right)$,
the signals that are distorted by the TX-IQI and RX-IQI, denoted by
$\underline{b}_{t}'\left(t\right)$ and $\underline{b}_{r}'\left(t\right)$,
respectively, can be expressed as follows{\small{}
\begin{eqnarray}
\underline{b}_{t}'\left(t\right)=\alpha_{t}\underline{b}\left(t\right)+\beta_{t}\underline{b}^{*}\left(t\right), &  & \underline{b}_{r}'\left(t\right)=\alpha_{r}\underline{b}\left(t\right)+\beta_{r}\underline{b}^{*}\left(t\right).\label{eq:txiqi}
\end{eqnarray}
}The parameters $(\alpha_{t},\beta_{t})$ and $(a_{r},\beta_{r})$
are the TX-IQI and RX-IQI parameters, respectively, which are defined
by

{\small{}
\begin{align}
\alpha_{t}=\cos(\frac{\phi_{t}}{2})+jg_{t}\sin(\frac{\phi_{t}}{2}),\thinspace\beta_{t}=g_{t}\cos(\frac{\phi_{t}}{2})-j\sin(\frac{\phi_{t}}{2}),\nonumber \\
\alpha_{r}=\cos(\frac{\phi_{r}}{2})+jg_{r}\sin(\frac{\phi_{r}}{2}),\thinspace\beta_{r}=g_{r}\cos(\frac{\phi_{r}}{2})-j\sin(\frac{\phi_{r}}{2}),
\end{align}
}where $g_{t}$, $\phi_{t}$ and $g_{r}$, $\phi_{r}$ are the amplitude
and the phase imbalance between the I and Q branches in the transmitter
and receiver ends, respectively. The transmit and receive amplitude
imbalances are often denoted in dB as $10\log(1+g_{t})$ and $10\log(1+g_{r})$,
respectively \cite{Tarighat2005}. The overall imbalance of a transceiver
is measured by the Image Rejection Ratio (IRR), which is defined by
$\mbox{IRR}_{t}(dB)\triangleq-10\log_{10}(\rho_{t})=-10\log_{10}(|{{\beta}_{t}}|^{2}/|{{\alpha}_{t}}|^{2})=-20\log_{10}(|{{\beta}_{t}}|/|{{\alpha}_{t}}|)$
and $\mbox{IRR}_{r}(dB)\triangleq-10\log_{10}(\rho_{r})=-20\log_{10}(|{{\beta}_{r}}|/|{{\alpha}_{r}}|)$,
where $\rho_{t}$ and $\rho_{r}$ could be equivalently viewed as
the normalized interference powers introduced by TX-IQI and RX-IQI,
respectively. 

In an OFDM system, discarding the samples corresponding to the $1^{st}$
and $\left(\frac{N}{2}+1\right)$-th subcarriers, the effect of the
IQI on the $n$-th subcarrier of the DSTBC-OFDM received signal is
basically introducing ICI from its $\left(N-n+2\right)$-th image
subcarrier \cite{Tarighat2007}. Based on the DSTBC-OFDM model in
Eq. (\ref{eq:freqDiff}), the frequency-domain received signal blocks
$\mathbf{Z'}_{k}\left(n\right)$ and $\mathbf{Z'}_{k+1}\left(n\right)$,
which are jointly distorted by TX-IQI and RX-IQI, are given by{\small{}
\begin{eqnarray}
\mathbf{\mathbf{Z'}}_{k}\left(n\right) & = & \left(\mathbf{A}_{t}\mathbf{A}_{r}\mathbf{\Lambda}(n)+\mathbf{B}_{t}^{*}\mathbf{B}_{r}\bar{\Lambda}(n)\right)\mathbf{S}_{k}\left(n\right)\nonumber \\
 &  & +\left(\mathbf{A}_{r}\mathbf{B}_{t}\mathbf{\Lambda}(n)+\mathbf{A}_{t}^{*}\mathbf{B}_{r}\mathbf{\bar{\Lambda}}(n)\right)\mathbf{\bar{S}}_{k}\left(n\right)\nonumber \\
 &  & +\mathbf{A}_{r}\mathbf{V}_{k}\left(n\right)+\mathbf{B}_{r}\mathbf{\bar{V}}_{k}\left(n\right),\label{eq:freDiff_k_IQ}
\end{eqnarray}
\begin{eqnarray}
\mathbf{\mathbf{Z'}}_{k+1}\left(n\right) & = & \left(\mathbf{A}_{t}\mathbf{A}_{r}\mathbf{\Lambda}(n)+\mathbf{B}_{t}^{*}\mathbf{B}_{r}\bar{\Lambda}(n)\right)\mathbf{S}_{k+1}\left(n\right)\nonumber \\
 &  & +\left(\mathbf{A}_{r}\mathbf{B}_{t}\mathbf{\Lambda}(n)+\mathbf{A}_{t}^{*}\mathbf{B}_{r}\mathbf{\bar{\Lambda}}(n)\right)\mathbf{\bar{S}}_{k+1}\left(n\right)\nonumber \\
 &  & +\mathbf{A}_{r}\mathbf{V}_{k+1}\left(n\right)+\mathbf{B}_{r}\mathbf{\bar{V}}_{k+1}\left(n\right)\nonumber \\
 & = & \left(\mathbf{A}_{t}\mathbf{A}_{r}\mathbf{\Lambda}(n)+\mathbf{B}_{t}^{*}\mathbf{B}_{r}\bar{\Lambda}(n)\right)\mathbf{S}_{k}(n)\mathbf{U}_{k+1}(n)\nonumber \\
 &  & +\left(\mathbf{A}_{r}\mathbf{B}_{t}\mathbf{\Lambda}(n)+\mathbf{A}_{t}^{*}\mathbf{B}_{r}\mathbf{\bar{\Lambda}}(n)\right)\mathbf{\bar{S}}_{k}(n)\mathbf{\bar{U}}_{k+1}(n)\nonumber \\
 &  & +\mathbf{A}_{r}\mathbf{V}_{k+1}\left(n\right)+\mathbf{B}_{r}\mathbf{\bar{V}}_{k+1}\left(n\right),\label{eq:eqfrediff1_k+1_IQ}
\end{eqnarray}
}where $\mathbf{\bar{\Lambda}}(n)=\mathbf{\Lambda}^{*}(N-n+2)$, $\mathbf{\bar{V}}_{k}\left(n\right)=\mathbf{V}_{k}^{*}\left(N-n+2\right)$,
$\mathbf{\bar{V}}_{k+1}\left(n\right)=\mathbf{V}_{k+1}^{*}\left(N-n+2\right)$,
$\mathbf{\bar{U}}_{k+1}(n)=\mathbf{U}_{k+1}^{*}(N-n+2)$ and $\mathbf{\bar{S}}_{k+1}(n)=\mathbf{\bar{S}}_{k}(n)\mathbf{\bar{U}}_{k+1}(n)=\mathbf{S}_{k+1}^{*}\left(N-n+2\right)$.
Moreover, the matrices $\mathbf{A}_{t/r}$, $\mathbf{B}_{t/r}$ are
constructed from the IQI parameters as follows
\begin{align}
\mathbf{A}_{t/r} & =\left[\begin{matrix}\alpha_{t/r} & 0\\
0 & \alpha_{t/r}^{*}
\end{matrix}\right],\;\;\mathbf{B}_{t/r}=\left[\begin{matrix}\beta_{t/r} & 0\\
0 & \beta_{t/r}^{*}
\end{matrix}\right].\label{eq:iqiparm}
\end{align}

\section{\label{sec:ber}Performance Analysis of DSTBC-OFDM under I/Q Imbalance}

In this section, we first analyze the impact of TX-IQI and RX-IQI
separately on an individual subcarrier in DSTBC-OFDM with M-PSK signaling,
and then present the combined effect of TX-IQI and RX IQI on the BER
performance. For notation simplicity, we omit the subcarrier index
$n$. In addition, recall that the entries of differentially-encoded
matrices, ${{\mathbf{S}}_{k}}$ and $\mathbf{{\mathbf{\bar{S}}}}_{k}$,
are sums of numerous products of PSK symbols, for a long input data
sequence, we apply the central limit theorem (CLT) to approximate
the distributions of the entries of ${{\mathbf{S}}_{k}}$, $\mathbf{S}_{k}^{H}\mathbf{\bar{S}}_{k+1}$
and $\mathbf{\bar{S}}_{k}^{H}\mathbf{S}_{k+1}$ by the uncorrelated
zero-mean Gaussian distributions ${{\mathbf{S}}_{k}}\sim\mathcal{N}\left(0,\,\frac{1}{2}\right)$,
$\mathbf{S}_{k}^{H}\mathbf{\bar{S}}_{k+1}\sim\mathcal{N}\left(0,\,\frac{1}{2}\right)$
and $\mathbf{\bar{S}}_{k}^{H}\mathbf{S}_{k+1}\sim N\left(0,\,\frac{1}{2}\right)$
with a variance of $\frac{1}{2}$ to satisfy the power constraint
{\small{}$\mathbf{S}_{k}\mathbf{S}_{k}^{H}=\mathbf{S}_{k}^{H}\mathbf{\bar{S}}_{k+1}\left(\mathbf{S}_{k}^{H}\mathbf{\bar{S}}_{k+1}\right)^{H}=\mathbf{\bar{S}}_{k}^{H}\mathbf{S}_{k+1}\left(\mathbf{\bar{S}}_{k}^{H}\mathbf{S}_{k+1}\right)^{H}=\mathbf{I}$}. 

On the other hand, after discarding the samples corresponding to the
$1^{st}$ and $\left(\frac{N}{2}+1\right)$-th subcarriers, we assume
that both the STBC block $\mathbf{S}_{k}^{H}(n)$ and the frequency-domain
channel response $\Lambda(n)$ of the desired subcarrier are independent
of their counterparts of the image subcarrier, namely, $\mathbf{\mathbf{\bar{S}}}_{k}^{H}(n)$
and $\bar{\Lambda}(n)$, respectively \cite{schenk2007performance}. 

To further simplify the notation, note that according to Eqs. \eqref{eq:freDiff_k_IQ}
and \eqref{eq:eqfrediff1_k+1_IQ}, in the presence of TX-IQI, for
each OFDM symbol, the equivalent transmitted signal $s'(k,n)=\alpha_{t}s(k,n)+\beta_{t}s^{*}\left(k,N-n+2\right)$
is a linear combination of two independent Gaussian random variables
whose phases are uniformly distributed. Thus, we can replace the IQI
parameters $\alpha_{t}$ and $\beta_{t}$ with their absolute values
$|\alpha_{t}|$ and $|\beta_{t}|$, respectively, in our following
performance analysis without changing the received signal statistical
properties. Similarly, from Eq. (\ref{eq:lmda}), the diagonal entries
of matrix $\mathbf{\Lambda}$ correspond to the DFT of the multi-path
CIR whose $L+1$ coefficients follow a zero-mean Gaussian distribution.
Thus, the non-zero entries of the diagonal matrix $\mathbf{\Lambda}$
follow a zero-mean Gaussian distribution with unit variance \cite{Torabi2007}.
Hence, the received signal also has a uniformly-distributed phase,
and we can also replace the RX-IQI parameters $\alpha_{r}$ and $\beta_{r}$
with their absolute values $|\alpha_{r}|$ and $|\beta_{r}|$ in our
following performance analysis. Thus, the IQI diagonal matrices $\mathbf{A}_{t}$,
$\mathbf{B}_{t}$, $\mathbf{A}_{r}$ and $\mathbf{B}_{r}$ become
$\mathbf{A}_{t/r}=\left|\alpha_{t/r}\right|\mathbf{I}$ and $\mathbf{B}_{t/r}=\left|\beta_{t/r}\right|\mathbf{I}$,
respectively. The accuracy of these assumptions is also supported
by the analysis of IQI in OFDM systems in \cite{schenk2007performance},
\cite{QiAissa2010}, where the impact of the IQI depends only on $\left|\alpha_{t/r}\right|$
and $\left|\beta_{t/r}\right|$.

For the following subsections, we quantify the asymptotic BER floor
caused by TX-IQI and RX-IQI and its corresponding equivalent signal-to-noise
ratio (SNR) compared to that of the IQI-free system. We also compare
the impact of IQI in differential and coherent STBC-OFDM. 

\subsection{DSTBC-OFDM under TX-IQI only\label{subsec:DSTBC-OFDM-under-TX-IQI}}

Based on the previous assumptions, the frequency-domain received signals
$\mathbf{\mathbf{Z'}}_{k}\left(n\right)$ and $\mathbf{\mathbf{Z'}}_{k+1}\left(n\right)$
in Eq. \eqref{eq:freDiff_k_IQ} and Eq. \eqref{eq:eqfrediff1_k+1_IQ},
in the presence of only TX-IQI can be expressed as{\small{}
\begin{equation}
\begin{array}{cl}
\mathbf{\mathbf{Z'}}_{TX,k} & =|{{a}_{t}}|\mathbf{\Lambda}\mathbf{S}_{k}+|{{\beta}_{t}}|\mathbf{\Lambda}\mathbf{\bar{S}}_{k}+\mathbf{V}_{k},\\
\mathbf{\mathbf{Z'}}_{TX,k+1} & =|{{a}_{t}}|\mathbf{\Lambda}\mathbf{S}_{k+1}+|{{\beta}_{t}}|\mathbf{\Lambda}\mathbf{\bar{S}}_{k+1}+\mathbf{V}_{k+1}\\
 & =|{{a}_{t}}|\mathbf{\Lambda}\mathbf{S}_{k}\mathbf{U}_{k+1}+|{{\beta}_{t}}|\mathbf{\Lambda}\mathbf{\bar{S}}_{k}\mathbf{\bar{U}}_{k+1}+\mathbf{V}_{k+1},
\end{array}
\end{equation}
}From Eq. (\ref{ML}), the decoding metric for the ML decoder under
TX-IQI is given by {\small{}
\begin{equation}
\begin{aligned}\mathbf{\mathbf{Z'}}_{TX,k}^{H}\mathbf{\mathbf{Z'}}_{TX,k+1} & =|\lambda|^{2}|a_{t}|^{2}\mathbf{U}_{k+1}+\mathbf{V}_{TX}\\
 & +|\lambda|^{2}\underset{\Theta_{TX}}{\underbrace{|\alpha_{t}\beta_{t}|\mathbf{S}_{k}^{H}\mathbf{\mathbf{\bar{S}}}_{k+1}+|\alpha_{t}\beta_{t}|\mathbf{\bar{S}}_{k}^{H}\mathbf{\mathbf{S}}_{k+1}}},
\end{aligned}
\label{eq:detectionmetric-t}
\end{equation}
}where $\mathbf{V}_{TX}=|\alpha_{t}|\mathbf{V}_{k}^{H}\mathbf{\Lambda\mathbf{S}}_{k}\mathbf{U}_{k+1}+|\alpha_{t}|\mathbf{S}_{k}^{H}\mathbf{\Lambda}^{H}\mathbf{V}_{k+1}$,
$|\lambda|^{2}\mathbf{I}=\mathbf{\Lambda}^{H}\mathbf{\Lambda}=(|\lambda_{1}^{2}(n)|+|\lambda_{2}^{2}(n)|)\mathbf{I}$.
Eq. \eqref{eq:detectionmetric-t} shows that TX-IQI will distort the
transmitted signal by introducing a Gaussian error term to the transmitted
signal. 

For a given channel realization at the desired subcarrier, the instantaneous
BER of the M-PSK symbols in $\mathbf{U}_{k+1}$ is determined by the
instantaneous equivalent signal-to-interference-plus-noise ratio (SINR)
of $\mathbf{U}_{k+1}$ in the decoding metric \cite{Torabi2007}.
Thus, we derive the average BER of DSTBC-OFDM by averaging the conditional
instantaneous BER over the probability distribution function (PDF)
of the equivalent SINR.

Asymptotically, it can be proved that when IRR is higher than a certain
level, TX-IQI will not cause any error floor when $\mathrm{SNR}\rightarrow\infty$
($i.e.$ ${{\sigma}^{2}}\rightarrow0$) due to the power constraint
on $\Theta_{TX}$. More specifically, let $A_{min}$ be the minimum
power of the error term that will cause an error in the detection
of the M-PSK constellation symbols. TX-IQI will not lead to a non-zero
asymptotic BER ($i.e.$ error floor) as $\mathrm{SNR}\rightarrow\infty$
unless $|\beta_{t}|/|\alpha_{t}|=\sqrt{\rho_{t}}<\sqrt{2}A_{min}/4$
(see Appendix \ref{subsec:Apa} for proof). For a normalized QPSK
signal with hard detection, it is easy to calculate that $A_{min,QPSK}=1/\sqrt{2}$,
which means that if $\mbox{IRR}_{t}(dB)=20\log(|{{\alpha}_{t}}|/|{{\beta}_{t}}|)$
is larger than $12.04$dB, there will no error floor for a DSTBC-OFDM
system using a QPSK constellation under TX-IQI as $\mathrm{SNR}\rightarrow\infty$.
For 8-PSK constellation, we have $A_{\min}=0.3827$, which means that
a BER floor appears only when IRR is smaller than $17.4$dB.

First, we derive the asymptotic BER when $\mathrm{SNR}\rightarrow\infty$
(BER floor) under severe TX-IQI, $i.e.$ $\mbox{IRR}_{t}(dB)<12.04(dB)$
for QPSK and $\mbox{IRR}_{t}(dB)<17.4(dB)$ for 8-PSK. In these cases,
since ${{\mathbf{S}}_{k}}^{H}\mathbf{{\mathbf{\bar{S}}}}_{k+1}$ and
${{\mathbf{\bar{S}}}_{k}}^{H}\mathbf{{\mathbf{S}}}_{k+1}$ are independent
Gaussian matrices, the detection metric in Eq. \eqref{eq:detectionmetric-t}
is given by
\begin{equation}
\mathbf{\mathbf{Z'}}_{TX,k}^{H}\mathbf{\mathbf{Z'}}_{TX,k+1}=|\lambda|^{2}\left(|{{a}_{t}}|^{2}\mathbf{U}_{k+1}+\Theta_{TX}\right),\label{eq:TX asmy}
\end{equation}
which is equivalent to transmitting a PSK symbol matrix $\mathbf{U}_{k+1}$
over an AWGN channel. Moreover, based on the assumed independence
between the desired subcarriers and their image subcarriers, $i.e.$
$E\{\mathbf{S}_{k}^{H}\mathbf{{\mathbf{\bar{S}}}}_{k+1}\mathbf{{\mathbf{S}}}_{k+1}^{H}{{\mathbf{\bar{S}}}_{k}}\}=0$,
each entry in $\Theta_{TX}$ is a zero-mean Gaussian random variable
whose variance is given by 
\begin{equation}
{{E}_{\Theta,TX}}=\frac{1}{4}E\{Tr({{\Theta_{TX}}^{H}}\Theta_{TX})\}=|{{\alpha}_{t}}{{\beta}_{t}}{{|}^{2}}.\label{eq:thetad}
\end{equation}
 The asymptotic instantaneous SINR ($i.e.$ ${{\sigma}^{2}}\rightarrow0$)
of the equivalent system in Eq. \eqref{eq:TX asmy} is
{\small{}\begin{eqnarray}
\eta_{TX}^{(a)} & = & Tr\{|a_{t}|^{4}\mathbf{U}_{k+1}^{H}\mathbf{U}_{k+1}\}/E\left\{ Tr\left(\Theta_{TX}^{H}\Theta_{TX}\right)\right\} \nonumber \\
 & = & \frac{|a_{t}|^{2}}{2|\beta_{t}|^{2}}=\frac{\rho_{t}^{-1}}{2}.
\end{eqnarray}}
Based on the general relationship between the BER and the instantaneous
SINR, denoted by $\eta_{t}^{(a)}$, of an M-PSK signal in \cite{Torabi2007},
the asymptotic BER (error floor) under severe TX-IQI, denoted by $P_{e,TX}^{(a)}$,
is given by
{\small{}\begin{equation}
	\begin{aligned}P_{e,Tx}^{(a)} & =\frac{1}{\log_{2}M}erfc\left(\sqrt{\eta_{t}^{(a)}}\sin\left(\pi/M\right)\right)\\
	& =\frac{1}{\log_{2}M}erfc\left(\sqrt{\frac{\rho_{t}^{-1}}{2}}\sin\left(\pi/M\right)\right).
	\end{aligned}
	\label{BER-psk-1}
	\end{equation}}

The BER floor given in Eq. \eqref{BER-psk-1} will only occur under
severe TX-IQI, which is $\mbox{IRR}_{t}(dB)<12.04(dB)$
for QPSK and ${\mbox{IRR}_{t}(dB)<17.4(dB)}$ for 8-PSK.
In other situations, there will be no BER floor caused by TX-IQI and
we will derive an equivalent SINR loss in these situations. Since
the detection of symbols in $\mathbf{U}_{k+1}$ is totally determined
by the detection metric in Eq. (\ref{eq:detectionmetric-t}), the
transmitted signal $\mathbf{U}_{k+1}$ could be viewed as being distorted
by the error vector $\Theta_{TX}=|{{\alpha}_{t}}{{\beta}_{t}}|{{\mathbf{S}}_{k}^{H}}\mathbf{{\mathbf{\bar{S}}}}_{k+1}+|{{\alpha}_{t}}{{\beta}_{t}}|{{\mathbf{\bar{S}}}_{k}^{H}}\mathbf{{\mathbf{S}}}_{k+1}$.
Based on the independence between the desired and image OFDM subcarriers,
each entry in $\Theta_{TX}$ is a zero-mean Gaussian random variable
whose variance is $|\alpha_{t}\beta_{t}|^{2}$ as given by Eq. \eqref{eq:thetad}.

\begin{figure}[tb]
\centering\includegraphics[width=3.5in]{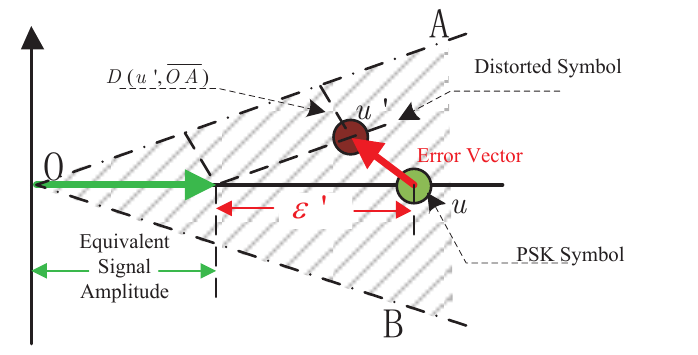}\caption{Equivalent signal amplitude in the presence of TX-IQI\label{fig: signal TX}}
\vspace{-0.5cm}
\end{figure}

Hence, to simplify the analysis, as depicted in Fig. \ref{fig: signal TX},
the PSK symbols in $\mathbf{U}_{k+1}$ can be viewed as being distorted
by a zero-mean complex Gaussian random vector $|\alpha_{t}\beta_{t}|n'$,
where $n'$ is a standard Gaussian random variable, $i.e.$ $E\{n'\}=0$
and $E\{|n'|^{2}\}=1$. The distance between the distorted symbol
$u'=u+n'$ and the PSK decision boundaries ($\overline{OA}$ and $\overline{OB}$
in Fig. \ref{fig: signal TX}), denoted as {\small{}$D(u',\overline{OA})$
}and {\small{}$D(u',\overline{OB})$} , respectively, are changed
by $n'$, which equivalently results in an amplitude loss $\varepsilon'$
in the signal amplitude. Further details can be found in Appendix
$\ref{subsec:Apb}$. According to the analysis in Appendix $\ref{subsec:Apb}$,
the equivalent transmitted signal power under the disturbance of $n'$
is given by
\begin{equation}
E_{s,TX}=|\alpha_{t}|^{4}(1+\sqrt{1+\cot(\pi/M)^{2}}\sqrt{|\rho_{t}|}n')^{2}/2.\label{eq:Estx}
\end{equation}

Clearly, the transmitted signal amplitude is affected by a zero-mean
random variable. Since the BER is determined by the tail of the error
function, the random variation in the transmitted signal power causes
more errors. To model the BER increase, we define the equivalent signal
amplitude loss due to TX-IQI as $\varepsilon_{TX}$. Since the BER
of M-PSK is approximately proportional to the inverse of the square
of the transmitted signal power \cite{Torabi2007}, thus $\varepsilon_{TX}$
is chosen to fit the expectation of the inverse of the square of the
transmitted power. Hence, using Eq. \eqref{eq:Estx}, we have{\small{}
\begin{equation}
E\left(\frac{1}{E_{s,TX}^{2}}\right)=\frac{4}{|a_{t}|^{8}}E\left(\frac{1}{(1+\sqrt{1+\cot(\pi/M)^{2}}\sqrt{\rho_{t}}n')^{4}}\right).
\end{equation}
}Thus, the equivalent amplitude loss $\varepsilon_{TX}$ should be
chosen to satisfy the following equation{\small{}
\begin{gather}
\frac{4}{|a_{t}|^{8}(1-\varepsilon_{TX})^{4}}=\frac{4}{|a_{t}|^{8}}E\left(\frac{1}{(1+\sqrt{1+\cot(\pi/M)^{2}}\sqrt{\rho_{t}}n')^{4}}\right).\label{eq:txeq}
\end{gather}
}The factor $4/|a_{t}|^{8}$ comes from the amplitude gain factor
$|a_{t}|^{2}/\sqrt{2}$ due to the transmission power normalization
and the presence of IQI. Since $\varepsilon_{TX}$ is a small real
number around $0$, by applying the Taylor expansion $(1+x)^{-4}\approx1-4x+10x^{2}$
to both sides of Eq. \eqref{eq:txeq} and ignoring the term corresponding
to $\varepsilon_{TX}^{2}$, we derive the approximated amplitude loss
due TX-IQI to be {\small{}
\begin{equation}
\begin{aligned} &  & 1+4\varepsilon_{TX}\approx & E\Bigl(1+4\sqrt{\rho_{t}\left(1+\cot(\pi/8)^{2}\right)}n'\\
 &  &  & +10(1+\cot(\pi/8)^{2})\rho_{t}|n'|^{2})\Bigr)\\
 &  & \longrightarrow\varepsilon_{TX}= & 2.5(1+\cot(\pi/M)^{2})\rho_{t}.
\end{aligned}
\label{eq:amploss}
\end{equation}
}Thus, the SINR loss introduced by TX-IQI is
\begin{eqnarray}
\eta_{loss}(dB) & = & -20\mbox{log}\left(|a_{t}|^{2}(1-\varepsilon_{TX})\right)\nonumber \\
 & \approx & 20\varepsilon_{TX}=50(1+\cot(\pi/M)^{2})\rho_{t}.\label{eq:gap}
\end{eqnarray}
The effective SINR under TX-IQI $\eta_{TX}$ becomes 
\begin{equation}
\eta_{TX}=\frac{|a_{t}|^{4}(1-\varepsilon_{TX})^{2}|\mathbf{\lambda}|^{2}}{4\sigma^{2}}.
\end{equation}
The equivalent instantaneous interference power is the expected power
of the entries of $\mathbf{\Theta}_{TX}$ for a given channel realization.
Since the frequency-domain channel coefficient $\mathbf{\mathbf{\lambda}}$
is a complex Gaussian variable, the term $|\mathbf{\lambda}|^{2}$
is a Chi-square random variable with 4 degrees of freedom. Hence,
the PDF of $\eta_{TX}$ can be written as 
\begin{equation}
p(\eta_{TX})=\frac{16\eta_{TX}\sigma^{4}}{|a_{t}|^{8}(1-\varepsilon_{TX})^{4}}\exp\left(\frac{-4\eta_{TX}\sigma^{2}}{|a_{t}|^{4}(1-\varepsilon_{TX})^{2}}\right).
\end{equation}
Thus, based on the BER expression in Eq. \eqref{BER-psk-1}, the BER
under TX-IQI, denoted by $P_{e,TX}$, can be averaged over the distribution
of the effective SINR $\eta_{TX}$ under TX-IQI, which is given by
{\small{}
\begin{equation}
P_{e,TX}=\int\limits _{0}^{\infty}\frac{1}{\log_{2}M}erfc\left(\sqrt{\eta_{TX}}\sin\left(\pi/M\right)\right)p(\eta_{TX})d\eta_{TX}.\label{eq:BER-TX-analytical}
\end{equation}
}Using the approximated BER expression in \cite{Torabi2007}, the
BER in Eq. \eqref{eq:BER-TX-analytical} has the following closed-form
{\small{}
\begin{equation}
\begin{aligned}P_{e,TX} & \approx\overset{_{\infty}}{\underset{_{0}}{\int}}0.2exp\left(-\frac{7\eta_{TX}}{2^{1.9\log_{2}M}+1}\right)p(\eta_{TX})d\eta_{TX}\\
 &  =0.2\left(1+\frac{1.75}{\sigma^{2}}\frac{|a_{t}|^{4}(1-\varepsilon_{TX})^{2}}{\left(1+M^{1.9}\right)}\right)^{-2}.
\end{aligned}
\label{eq:BER-TX-closedform}
\end{equation}
}{\small \par}

Note that with moderate TX-IQI, the interference due to TX-IQI leads
to an equivalent signal power degradation in the transmitted signal,
which means that its impact does not change with the noise power,
resulting in a constant SINR gap between the BER curve of an ideal
system and that of a TX-IQI-distorted system. This SINR gap is roughly
$-20\log(1-\varepsilon_{TX})$ in dB according to our analysis in
Eq. \eqref{eq:gap}. In addition, the signal amplitude loss $\varepsilon_{TX}$
shown in Eq. \eqref{eq:amploss} is determined by both TX-IQI and
the size of the signal constellation. As shown in Eq. \eqref{eq:gap},
under the same TX-IQI level (\emph{i.e.} same $|\rho_{t}|$), the
SNR loss due to TX-IQI of 8-PSK is roughly $\frac{1+\cot(\pi/8)^{2}}{1+\cot(\pi/4)^{2}}=3.4$
times of that of QPSK (in dB), which confirms that 8-PSK is much more
sensitive to TX-IQI than QPSK. 

\subsection{DSTBC-OFDM under RX-IQI only\label{subsec:DSTBC-OFDM-under-RX-IQI}}

Similar to the case of TX-IQI, the frequency-domain received signal
in the presence of RX-IQI, denoted by $\mathbf{\mathbf{Z'}}_{RX,k}\left(n\right)$
and $\mathbf{\mathbf{Z'}}_{RX,k+1}\left(n\right)$, is given by{\small{}
\begin{align}
\mathbf{\mathbf{Z'}}_{RX,k} & =|\alpha_{r}|\mathbf{\Lambda}{{\mathbf{S}}_{k}}+|{{\beta}_{r}}|\mathbf{\mathbf{\bar{\Lambda}}}{{\mathbf{\bar{S}}}_{k}}+|{{\alpha}_{r}}|{{\mathbf{V}}_{k}}+|{{\beta}_{r}}|{{\mathbf{\bar{V}}}_{k}},\nonumber \\
\mathbf{\mathbf{Z'}}_{RX,k+1} & =|{{\alpha}_{r}}|\mathbf{\Lambda}{{\mathbf{S}}_{k}}\mathbf{U}_{k+1}+|{{\beta}_{r}}|\mathbf{\mathbf{\bar{\Lambda}}}{{\mathbf{\bar{S}}}_{k}}\mathbf{\mathbf{\bar{U}}}_{k+1}\nonumber \\
 & +|{{\alpha}_{r}}|\mathbf{V}_{k+1}+|{{\beta}_{r}}|\mathbf{\bar{V}}_{k+1}.
\end{align}
}From Eq. (\ref{ML}), the decoding metric for the ML decoder under
RX-IQI becomes {\small{}
\begin{equation}
\begin{aligned}\mathbf{\mathbf{Z'}}_{RX,k}^{H}\mathbf{\mathbf{Z'}}_{RX,k+1} & =|\lambda|^{2}|\alpha_{r}{{|}^{2}}\mathbf{U}_{k+1}+{{\mathbf{V}}_{RX}}\\
 & +\underbrace{|{{\alpha}_{r}}{{\beta}_{r}}|\left({{\mathbf{S}}_{k}}^{H}{{\mathbf{\Lambda}}^{H}}\mathbf{\bar{\Lambda}{\mathbf{\bar{S}}}}_{k+1}+{{\mathbf{\bar{S}}}_{k}}^{H}{{\mathbf{\bar{\Lambda}}}^{H}}\mathbf{\Lambda{\mathbf{S}}}_{k+1}\right)}_{\Theta_{RX}},
\end{aligned}
\label{eq:detectionmetric}
\end{equation}
}where${{\mathbf{V}}_{RX}}=|{{\alpha}_{r}}{{|}^{2}}{{\mathbf{V}}_{k}}^{H}\mathbf{\Lambda{\mathbf{S}}}_{k}\mathbf{U}_{k+1}+|{{\alpha}_{r}}{{|}^{2}}{{\mathbf{S}}_{k}}^{H}{{\mathbf{\Lambda}}^{H}}\mathbf{V}_{k+1}$. 

The detection of symbols in $\mathbf{U}_{k+1}$ under RX-IQI is determined
by the detection metric in Eq. (\ref{eq:detectionmetric}). Unlike
TX-IQI, the interference term is determined not only by the channel
of the desired subcarrier $\mathbf{\Lambda}$, but also by the channel
of the image subcarrier $\mathbf{\bar{\Lambda}}$. In addition, due
to our independent assumption between the desired subcarrier and its
image subcarrier, we have $E\{{{\mathbf{S}}_{k}}^{H}{{\mathbf{\Lambda}}^{H}}\bar{\Lambda}\mathbf{{\mathbf{\bar{S}}}}_{k+1}{{\mathbf{S}}_{k+1}}^{H}{{\mathbf{\Lambda}}^{H}}\bar{\Lambda}\mathbf{{\mathbf{\bar{S}}}}_{k}\}=\mathbf{0}$.
Thus, the conditional average power of each entry of the matrix $\mathbf{\Theta}_{RX}$
is given by
\begin{equation}
{{E}_{\Theta,RX}}=\frac{1}{4}E\{Tr({{\mathbf{\Theta}_{RX}}^{H}}\mathbf{\Theta}_{RX}\left|\mathbf{\Lambda},\mathbf{\bar{\Lambda}}\right.\}=|{{\alpha}_{r}}{{\beta}_{r}}{{|}^{2}}|{\mathbf{\lambda}}{{|}^{2}}|\mathbf{\bar{\lambda}}{{|}^{2}}.
\end{equation}

Similarly, the conditional average signal power $E_{S}$ and noise
power ${E}_{v}$ for a given channel realization can be expressed
as follows 
\begin{align}
{\normalcolor {\normalcolor {\color{red}{\normalcolor E_{S}}}}} & {\normalcolor {\normalcolor {\color{red}{\normalcolor =\frac{|\lambda|^{4}|{{\alpha}_{r}}{{|}^{4}}}{4}}{\normalcolor E\left\{ Tr\left(\mathbf{U}_{k+1}^{H}\mathbf{U}_{k+1}\right)\right\} =\frac{1}{2}|{{\alpha}_{r}}\lambda{{|}^{4}}}}}},\nonumber \\
{{E}_{v}} & =\frac{1}{4}E\left\{ Tr({\mathbf{V}}_{r}^{H}{\mathbf{V}}_{r})\left|\mathbf{\Lambda},\mathbf{\bar{\Lambda}}\right.\right\} =2|{{\alpha}_{r}}{{|}^{4}}|\lambda{{|}^{2}}{{\sigma}^{2}}.
\end{align}
Therefore, the conditional instantaneous SINR $\eta_{r}$ of $\mathbf{U}_{k+1}$
in the differential decoding metric for a given channel realization
is given by 
\begin{equation}
{\eta_{RX}}=\frac{{E_{S}}}{{{E}_{\Theta,RX}}+{{E}_{v}}}=\frac{|\lambda{{|}^{2}}}{2|\bar{\lambda}{{|}^{2}}\rho_{r}+4{{\sigma}^{2}}},\label{eq:etaprecise}
\end{equation}
where $\rho_{r}=|\beta_{r}|^{2}/|\alpha_{r}|^{2}$. To start with,
we analyze the asymptotic performance when $\mathrm{SNR}\rightarrow\infty$.
By setting ${\sigma}^{2}\rightarrow0$ , the asymptotic equivalent
SINR becomes
\begin{equation}
{{\eta}_{RX}^{\left(a\right)}}=\underset{\sigma^{2}\rightarrow0}{\mbox{lim}}\:\eta_{RX}=\frac{|\lambda{{|}^{2}}}{2|\mathbf{\mathbf{\bar{\lambda}}}{{|}^{2}}\rho_{r}}.\label{etaasym}
\end{equation}
Since $\lambda$ and $\mathbf{\bar{\lambda}}$ are independent complex
Gaussian random variables, the ratio of their squared-absolute values,
denoted by $X=\left|\lambda\right|^{2}/|\overline{\lambda}|^{2}$,
follows the F-distribution \cite{degroot2010} with a probability
density function given by $p(X)=F(x,4,4)$, where $F(x,b,c)=I_{\frac{bx}{bx+c}}\left(\frac{b}{2},\frac{c}{2}\right)$
and $I$ is the regularized incomplete beta function. It can be proved
that $E\left(X\right)=\underset{x}{\int}X\,p(X)\,dX=2$. Hence, the
asymptotic average equivalent SINR is given by{\small{}
\begin{align}
E\left(\eta_{RX}^{\left(a\right)}\right) & =\frac{\left|\alpha_{r}\right|^{2}}{2\left|\beta_{r}\right|^{2}}E\left(\frac{\left|\lambda\right|^{2}}{\left|\overline{\lambda}\right|^{2}}\right)=\frac{\left|\alpha_{r}\right|^{2}}{2\left|\beta_{r}\right|^{2}}E\left(X\right)=\frac{\left|\alpha_{r}\right|^{2}}{\left|\beta_{r}\right|^{2}}\nonumber \\
 & =1/\rho_{r}=\mathrm{IRR}{}_{r}.\label{eeta}
\end{align}
}Based on the general relationship between the BER and the instantaneous
SINR $\eta$ of an M-PSK signal in \cite{Torabi2007}, the average
asymptotic BER (error floor) in the presence of RX-IQI, denoted by
$P_{e,RX}^{(a)}$, is given by{\small{}
\begin{equation}
P_{e,RX}^{(a)}=\int\limits _{0}^{\infty}{\frac{1}{{{\log}_{2}}M}erfc\left(\sqrt{{\eta_{RX}^{\left(a\right)}}}\sin\left(\pi/M\right)\right)p({\eta_{RX}^{\left(a\right)}})d{\eta_{RX}^{\left(a\right)}}},\label{BER-RX-analytical}
\end{equation}
}where $p\left({\eta_{RX}^{\left(a\right)}}\right)=2\rho_{r}F(2\rho_{r}x,4,4)$
is the probability distribution function of SINR $\eta_{RX}^{\left(a\right)}$
given in Eq. (\ref{etaasym}).\vspace{-0.1em}

We note that $\beta_{r}\ll\alpha_{r}$ since the interference power
from the image subcarrier is much smaller than that of the desired
subcarrier. Hence, the interference due to RX-IQI can be approximated
by a Gaussian distribution and incorporated into the noise term \cite{Jafar}
without loss of generality. Since $|\bar{\lambda}|^{2}$ is the sum
of two independent and identically distributed (i.i.d) zero-mean complex
Gaussian random variables with unit variance, and it is given by $|\bar{\lambda}|^{2}={|{\lambda_{1}}}(N-n+2)|^{2}+{|{\lambda}_{2}}(N-n+2)|^{2}$,
its average power is $E\{2|\bar{\lambda}{{|}^{2}}\rho_{r}\}=4\rho_{r}$.
Thus, the instantaneous SINR $\eta_{RX}$ in Eq. (\ref{eq:etaprecise})
becomes a Chi-square random variable with $4$ degrees of freedom
given by{\small{}
\begin{equation}
\eta_{RX}\left|_{\beta_{r}\ll\alpha_{r}}\right.=\frac{1}{4}\left|\lambda\right|^{2}\left(\rho_{r}+\sigma^{2}\right)^{-1}.\label{eq:etanonasym}
\end{equation}
}The distribution of $\eta_{RX}$ given by
\begin{equation}
p(\eta_{RX})=16\eta_{RX}(\rho_{r}+\sigma^{2})^{2}\exp\left(-4\eta_{RX}(\rho_{r}+\sigma^{2})\right).\label{eq:pEtaRX}
\end{equation}
From Eq. (\ref{eq:etanonasym}), the BER floor appears roughly at
the SNR level where the RX-IQI interference power $\rho_{r}$, overwhelms
the noise power $\sigma^{2}$ (we assume 10 times larger), which means
that the BER floor approximately appears when the corresponding SINR,
denoted as ${{\eta}_{\mathrm{floor}}}$, satisfies the following conditions{\small{}
\begin{equation}
\begin{aligned} & {{\eta}_{\mathrm{floor}}}\gg1/\rho_{r}\\
\rightarrow & {{\eta}_{\mathrm{floor}}}(\mathrm{dB})\approx\mathrm{IRR(dB)+10dB}.
\end{aligned}
\label{eq:flooraper}
\end{equation}
}Let ${\eta}_{\mathrm{ideal}}$ be the equivalent SNR of an IQI-free
DSTBC-OFDM system that has a BER equal to the BER floor $P_{e,RX}^{(a)}$,
which is given by{\small{}
\begin{equation}
{\eta}_{\mathrm{ideal}}(\mathrm{dB})=-10\log_{10}(\rho_{r})=\mathrm{IRR}_{r}(\mathrm{dB}).\label{eq:flooreqideal}
\end{equation}
}This indicates that the best BER under RX-IQI equals the BER of an
IQI-free system when SNR is equal to IRR. The BER at any SNR under
RX-IQI can be calculated by using $\eta_{RX}$ and its PDF $p(\eta_{RX})$
in Eqs. (\ref{eq:etanonasym}) and (\ref{eq:pEtaRX})
when evaluating the integral in Eq. (\ref{BER-RX-analytical}). Similar
to Eq. \eqref{eq:BER-TX-closedform}, this integral could also be
approximated by a closed-form as follows{\small{}
\begin{align}
P_{e,RX} & \approx0.2\left(1+1.75\frac{(\rho_{r}+\sigma^{2})^{-1}}{M^{1.9}+1}\right)^{-2} \nonumber \\
&=0.2\left(1+1.75\frac{\mathrm{SNR_{eq}}}{M^{1.9}+1}\right)^{-2}.\label{eq:BER-RX-closedform}
\end{align}
}The term $(\rho_{r}+{\sigma}^{2})^{-1}$ can be viewed as an equivalent
SNR, denoted by $\mathrm{SNR_{eq}}$ ($\mbox{\ensuremath{\mathrm{SNR_{eq}}}= \ensuremath{\mbox{SNR}^{-1}} + \ensuremath{\mbox{IRR}^{-1}}}$),
which is the harmonic mean of the SNR ($\mbox{\ensuremath{{\sigma}^{2}}}=\mbox{SNR}^{-1}$)
and the $\mathrm{IRR}_{r}$ $\left(\mbox{\ensuremath{\rho_{r}}}=\mbox{IR\ensuremath{\mbox{R}^{-1}}}\right)$
and is always less than the minimum of the two.

For the high SNR scenario, in the case of high RX-IQI levels and hence
a low IRR level, the equivalent SNR $\mathrm{SNR_{eq}}$ and hence
the BER $P_{e,RX}$ in Eq. \eqref{eq:BER-RX-closedform} will be dominated
by the IRR level, $i.e.$ $\mbox{\ensuremath{\mathrm{SNR_{eq}}}}\approx\mbox{IRR\ensuremath{_{r}} \ensuremath{\approx}\ 1/\ensuremath{\rho_{r}}}$.
Moreover, the BER is more sensitive to both noise and RX-IQI effects
for a higher-order signal constellation $i.e.$ larger $M$. When
the noise is negligibly small, the $\mathrm{SNR_{eq}}$ of QPSK modulation
is $\frac{8^{1.9}+1}{4^{1.9}+1}=3.55$ times larger than that of 8PSK
modulation, which indicates that the IRR gap for BER floor between
QPSK and 8PSK is $10\mbox{log}(36.55)=5.5\mathrm{dB}$. On the other
hand, for the IQI-free scenario, the BER $P_{e,RX}$ in Eq. \eqref{eq:BER-RX-closedform}
will be dominated by the SNR level, $i.e.$ $\mbox{SN\ensuremath{R_{eq}}}\approx\mbox{SNR \ensuremath{\approx}\ 1/\ensuremath{\sigma}}^{2}$.
As SNR increases, Eq. \eqref{eq:BER-RX-closedform} clearly shows
that the diversity order is two as expected. 

\subsection{DSTBC-OFDM under joint TX-IQI and RX-IQI}

Following our assumptions in seperate TX-IQI and RX-IQI performance
analysis, the frequency-domain received signal in the presence of
both TX-IQI and RX-IQI, denoted by $\mathbf{\mathbf{Z'}}_{TR,k}\left(n\right)$
and $\mathbf{\mathbf{Z'}}_{TR,k+1}\left(n\right)$, is given by{\small{}
\begin{eqnarray}
\mathbf{\mathbf{Z'}}_{TR,k} & = & \left(|\alpha_{t}\alpha_{r}|\mathbf{\Lambda}+|\beta_{t}\beta_{r}|\bar{\Lambda}\right)\mathbf{S}_{k}+\left(|\alpha_{r}\beta_{t}|\mathbf{\Lambda}+|\alpha_{t}\beta_{r}|\mathbf{\bar{\Lambda}}\right)\mathbf{\bar{S}}_{k}\nonumber \\
 &  & +|\alpha_{r}|\mathbf{V}_{k}+|\beta_{r}|\mathbf{\bar{V}}_{k},\label{eq:freDiff_k_IQ-1}
\end{eqnarray}
\begin{eqnarray}
\mathbf{\mathbf{Z'}}_{TR,k+1}\left(n\right) & = & \left(|\alpha_{t}\alpha_{r}|\mathbf{\Lambda}+|\beta_{t}\beta_{r}|\bar{\Lambda}\right)\mathbf{S}_{k}\mathbf{U}_{k+1}\nonumber \\
 &  & +\left(|\alpha_{r}\beta_{t}|\mathbf{\Lambda}+|\alpha_{t}\beta_{r}|\mathbf{\bar{\Lambda}}\right)\mathbf{\bar{S}}_{k}\mathbf{\bar{U}}_{k+1}.\nonumber \\
 &  & +|\alpha_{r}|\mathbf{V}_{k+1}+|\beta_{r}|\mathbf{\bar{V}}_{k+1}\label{eq:eqfrediff1_k+1_IQ-1}
\end{eqnarray}
}{\small \par}

After ignoring the small terms that contain high-order terms of $|\beta_{t}|$
or $|\beta_{r}|$, the decoding metric could be simplified as{\small{}
\begin{equation}
\begin{aligned}\mathbf{\mathbf{Z'}}_{TR,k}^{H}\mathbf{\mathbf{Z'}}_{TR,k+1} & =|\lambda|^{2}\left(|\alpha_{t}\alpha_{r}|^{2}\mathbf{U}_{k+1}\right)+\mathbf{V}_{TR}\\
 & +\underset{\Theta_{TX}^{'}}{|\lambda|^{2}\underbrace{|\alpha_{r}{}^{2}\alpha_{t}\beta_{t}|\mathbf{S}_{k}^{H}\mathbf{\mathbf{\bar{S}}}_{k+1}+|\alpha_{r}{}^{2}\alpha_{t}\beta_{t}|\mathbf{\bar{S}}_{k}^{H}\mathbf{\mathbf{S}}_{k+1}}}\\
 & +\underset{\Theta_{RX}^{'}}{\underbrace{|\alpha_{t}{}^{2}\alpha_{r}\beta_{r}|\left({{\mathbf{S}}_{k}}^{H}{{\mathbf{\Lambda}}^{H}}\mathbf{\bar{\Lambda}{\mathbf{\bar{S}}}}_{k+1}+{{\mathbf{\bar{S}}}_{k}}^{H}{{\mathbf{\bar{\Lambda}}}^{H}}\mathbf{\Lambda{\mathbf{S}}}_{k+1}\right)}},
\end{aligned}
\end{equation}
}where {\small{}$\mathbf{V}_{TR}=|\alpha_{t}\alpha_{r}^{2}|{\mathbf{V}}_{k}^{H}\mathbf{\Lambda{\mathbf{S}}}_{k}\mathbf{U}_{k+1}+|\alpha_{t}\alpha_{r}^{2}|\mathbf{S}_{k}^{H}\mathbf{\Lambda}^{H}\mathbf{V}_{k+1}$.}{\small \par}

Since $\Theta_{TX}$ and $\Theta_{RX}$ are uncorrelated zero-mean
random variables, the interference mechanism under joint TX-IQI and
RX-IQI is the direct combination of each distortion. By applying a
similar analysis to what we did in Subsections \ref{subsec:DSTBC-OFDM-under-TX-IQI}
and \ref{subsec:DSTBC-OFDM-under-RX-IQI}, the instantaneous SINR
in the presence of both TX-IQI and RX-IQI is given by{\small{}
\begin{align}
\eta_{TR} & =\frac{|\alpha_{t}\alpha_{r}|^{4}(1-\varepsilon_{TX})^{2}|\mathbf{\lambda}|^{2}}{4(|\alpha_{t}\alpha_{r}^{2}|^{2}\sigma^{2}+|\alpha_{t}{}^{2}\alpha_{r}^{2}|^{2}\rho_{r})}\nonumber \\
 & =\frac{(1-\varepsilon_{TX})^{2}|\mathbf{\lambda}|^{2}}{4(\sigma^{2}/|\alpha_{t}|^{2}+\rho_{r})}\nonumber \\
 & \thickapprox\frac{(1-\varepsilon_{TX})^{2}|\mathbf{\lambda}|^{2}}{4(\sigma^{2}+\rho_{r})}.
\end{align}
}{\small \par}

Similarly, $\eta_{TR}$ is a Chi-square random variable with 4 degrees
of freedom and the BER could be straightforwardly derived by replacing
the noise power term in Eq. \eqref{eq:BER-TX-analytical} and \eqref{eq:BER-TX-closedform}
as $\sigma^{2}\rightarrow(\sigma^{2}+\rho_{r})$. Note that since
RX-IQI causes much more BER degradation than TX-IQI, the combined
effect of TX-IQI and RX-IQI is mainly determined by the level of RX-IQI.
However, this does not mean compensation of TX-IQI is less important
than that of RX-IQI, because as we will discuss in the next section,
TX-IQI degrades the estimation accuracy of RX-IQI.

\subsection{Comparison with Coherent Detection\label{cohcompare}}

In this subsection, we compare the effect of TX-IQI and RX-IQI in
differential detection with that in coherent detection. For coherent
detection, the information block $\mathbf{\mathbf{U}}_{k+1}\text{(}n\text{)}$
is directly transmitted (we remove the index $k+1$ and $n$ for notational
simplicity). The received signal block under the TX-IQI, denoted by
$\mathbf{\mathbf{Z'}}_{TX,coh},$ and that under the RX-IQI, denoted
by $\mathbf{\mathbf{Z'}}_{RX,coh}$, become{\small{}
\begin{equation}
\begin{aligned}\mathbf{\mathbf{Z'}}_{TX,coh} & =|\alpha_{t}|\mathbf{\Lambda}\mathbf{\mathbf{U}}\text{+}|\alpha_{t}|\mathbf{\Lambda}\mathbf{\bar{U}}+\mathbf{V},\\
\mathbf{\mathbf{Z'}}_{RX,coh} & =|a_{r}|\mathbf{\Lambda}\mathbf{\mathbf{U}}\text{+}|\beta_{r}|\mathbf{\bar{\Lambda}}\mathbf{\bar{U}}+|a_{r}|\mathbf{V}+|\beta_{r}|\mathbf{\bar{V}}.
\end{aligned}
\end{equation}
}{\small \par}

Assuming that the receiver has perfect CSI, the coherent detection
process at the receiver can be expressed as \cite{Alamouti1998}
\begin{eqnarray}
\mathbf{\hat{U}}_{t}(n) & = & \underset{\mathbf{U}}{\mathop{\arg\max}}\,\left\{ {{\mathbf{U}}^{H}}\mathbf{\Lambda}^{H}\mathbf{\mathbf{Z'}}_{TX,coh}\right\} ,\nonumber \\
\mathbf{\hat{U}}_{r}(n) & = & \underset{\mathbf{U}}{\mathop{\arg\max}}\,\left\{ {{\mathbf{U}}^{H}}\mathbf{\Lambda}^{H}\mathbf{\mathbf{Z'}}_{RX,coh}\right\} ,\label{eq:Ucoh}
\end{eqnarray}
where $\mathbf{\Lambda}{}^{H}{\mathbf{\mathbf{Z}'}_{TX,coh}}$ and
$\mathbf{\Lambda}{}^{H}{\mathbf{\mathbf{Z}'}_{RX,coh}}$ can be approximated
as follows{\small{}
\begin{equation}
\begin{aligned}\mathbf{\Lambda}{}^{H}{\mathbf{\mathbf{Z}'}_{TX,coh}}\approx & |\mathbf{\lambda}{{|}^{2}}(|{{\alpha}_{t}}|\mathbf{U}+\underset{\Theta_{TX}^{c}}{\underbrace{|{{\beta}_{t}}|\mathbf{\bar{U}}}})+\mathbf{\Lambda}{}^{H}{\mathbf{V}},\end{aligned}
\label{eq:tcoh}
\end{equation}
}{\small \par}

{\small{}
\begin{equation}
\begin{aligned}\mathbf{\Lambda}{}^{H}{\mathbf{\mathbf{Z}'}_{RX,coh}}\approx & |{{\alpha}_{r}}||\mathbf{\lambda}{{|}^{2}}\mathbf{U}+\underset{\Theta_{RX}^{c}}{\underbrace{|{{\beta}_{r}}|\mathbf{\Lambda}{}^{H}\mathbf{\bar{\Lambda}}\mathbf{\bar{U}}}}+\mathbf{\Lambda}{}^{H}\mathbf{A}_{r}\mathbf{V}.\end{aligned}
\label{eq:rcoh}
\end{equation}
}{\small \par}

Comparing Eq. \eqref{eq:tcoh} and Eq. \eqref{eq:rcoh} with the detection
metric in Eq. \eqref{eq:detectionmetric-t} and Eq. \eqref{eq:detectionmetric},
for the case of IQI-free system ($i.e.$ ${\alpha}_{t,r}=1$ and ${\beta}_{t,r}=0$),
the noise power is half of its value in differential detection, which
leads to a 3dB loss in SNR as observed in \cite{Hughes2000}.

In the case of TX-IQI, the power of the error term caused by TX-IQI,
denoted by $E_{\Theta,TX}^{c}$, normalized with respect to the signal
power $|{\alpha}_{t}|^{2}$ in coherent detection under TX-IQI, is
given by{\small{}
\begin{eqnarray}
\frac{{E_{\Theta,TX}^{c}}}{|{\alpha}_{t}|^{2}} & = & \frac{1}{4}E\{Tr({{\left(\Theta_{TX}^{c}\right)}^{H}}\Theta_{TX}^{c})\}\nonumber \\
 & = & \left|\frac{{\beta}_{t}}{{2\alpha}_{t}}\right|^{2}Tr(\left(\mathbf{\bar{U}}\text{(}n\text{)}\right)^{H}\mathbf{\bar{U}}\text{(}n\text{)})=\frac{1}{2}\left|\frac{{\beta}_{t}}{{\alpha}_{t}}\right|^{2}.\label{eq:txcohpower}
\end{eqnarray}
}{\small \par}

Eq. \eqref{eq:txcohpower} is half of the normalized power of the
TX-IQI error term in the differential case, $i.e.$ $E_{\Theta,TX}/|{\alpha}_{t}|^{4}=\left|\frac{{\beta}_{t}}{{\alpha}_{t}}\right|^{2}$,
where $E_{\Theta,TX}$ is given in Eq. \eqref{eq:thetad}. Similarly,
following the RX-IQI analysis in Subsection \ref{subsec:DSTBC-OFDM-under-RX-IQI},
the conditional SINR $\eta_{RX,c}$ for the coherent detection of
$\mathbf{U}\text{(}n\text{)}$ for a given channel realization under
RX-IQI is given by 
\begin{equation}
{{\eta}_{RX,c}}=\frac{|\lambda{{|}^{2}}}{|\bar{\lambda}{{|}^{2}}\rho_{r}+2{{\sigma}^{2}}},
\end{equation}
whose power is also half of its value in differential detection per
Eq. \eqref{eq:etaprecise}. The reason for the halved interference
power in the coherent case is that, in both the TX-IQI and RX-IQI
cases, a doubled interference power will be introduced to the detection
SINR because both the previous block and current block are affected
by interference due to TX-IQI or RX-IQI in differential detection,
while in coherent detection we assume perfect CSI of the IQI-free
system. Thus, the BER of coherent detection can be obtained by setting
both the noise power $\sigma^{2}$ and the IQI interference power
$\rho_{t}$ and $\rho_{r}$ in the BER expression of differential
detection as $\sigma^{2}\rightarrow\sigma_{c}^{2}/2$, $\rho_{t}\rightarrow\rho_{t}^{(c)}/2$
and $\rho_{r}\rightarrow\rho_{r}^{(c)}/2$, where $\sigma_{c}^{2}$,
$\rho_{t}^{(c)}$ and $\rho_{r}^{(c)}$ are the noise power, TX-IQI
and RX-IQI interference power of the coherent detection system, respectively.
Equivalently, we can say that under the same noise power and transceiver
IQI levels, the performance gap between differential and coherent
STBC detection consists of a 3dB loss in SNR and also a 3dB loss in
IRR in differential detection. 

\section{\label{sec:compensateion}IQI Estimation and Compensation Algorithm
in DSTBC-OFDM}

\subsection{Widely-linear (WL) Compensation\label{subsec:Widely-linear-Compensation}}

The frequency-domain IQI-distorted received signals in Eqs. (\ref{eq:freDiff_k_IQ})
and (\ref{eq:eqfrediff1_k+1_IQ}) can be expressed in the widely-linear
equivalent form as follows 
{\small{}
\begin{equation}
\left[ \begin{matrix}
{{{\mathbf{{Z}'}}}_{k}}\left( n \right)  \\
{{{\mathbf{{\bar{Z}}'}}}_{k}}\left( n \right)  \\
\end{matrix} \right]=\mathbf{\Phi }(n)\left[ \begin{matrix}
{{\mathbf{S}}_{k}}\text{(}n\text{)}  \\
{{{\mathbf{\bar{S}}}}_{k}}\text{(}n\text{)}  \\
\end{matrix} \right]+\left[ \begin{matrix}
{{\mathbf{A}}_{r}}{{\mathbf{V}}_{k}}(n)\text{+}{{\mathbf{B}}_{r}}{{{\mathbf{\bar{V}}}}_{k}}(n)  \\
\mathbf{A}_{r}^{*}{{{\mathbf{\bar{V}}}}_{k}}(n)\text{+}\mathbf{B}_{r}^{*}{{\mathbf{V}}_{k}}(n)  \\
\end{matrix} \right],\label{WLTRIQ-2}
\end{equation}
}{\small \par}

{\small{}
\begin{equation}
\left[ \begin{matrix}
{{{\mathbf{{Z}'}}}_{k+1}}\left( n \right)  \\
{{{\mathbf{{\bar{Z}}'}}}_{k+1}}\left( n \right)  \\
\end{matrix} \right]=\mathbf{\Phi }(n)\left[ \begin{matrix}
{{\mathbf{S}}_{k+1}}\text{(}n\text{)}  \\
{{{\mathbf{\bar{S}}}}_{k+1}}\text{(}n\text{)}  \\
\end{matrix} \right]+\left[ \begin{matrix}
{{\mathbf{A}}_{r}}{{\mathbf{V}}_{k+1}}(n)\text{+}{{\mathbf{B}}_{r}}{{{\mathbf{\bar{V}}}}_{k+1}}(n)  \\
\mathbf{A}_{r}^{*}{{{\mathbf{\bar{V}}}}_{k+1}}(n)\text{+}\mathbf{B}_{r}^{*}{{\mathbf{V}}_{k+1}}(n)  \\
\end{matrix} \right],
\label{WLTRIQ-1-1}
\end{equation}
}{\small}
where $\mathbf{S}_{k+1}\text{(}n\text{)}=\mathbf{S}_{k}(n)\mathbf{U}_{k+1}(n)$ and
\begin{equation}
\mathbf{\Phi }(n)=\left[ \begin{matrix}
{{\mathbf{A}}_{r}}{{\mathbf{A}}_{t}}\mathbf{\Lambda }(n)\text{+}{{\mathbf{B}}_{r}}\mathbf{B}_{t}^{*}\mathbf{\bar{\Lambda }}(n) & {{\mathbf{A}}_{r}}{{\mathbf{B}}_{t}}\mathbf{\Lambda }(n)\text{+}\mathbf{A}_{t}^{*}{{\mathbf{B}}_{r}}\mathbf{\bar{\Lambda }}(n)  \\
\mathbf{A}_{r}^{*}\mathbf{B}_{t}^{*}\mathbf{\bar{\Lambda }}(n)\text{+}{{\mathbf{A}}_{t}}\mathbf{B}_{r}^{*}\mathbf{\Lambda }(n) & \mathbf{A}_{r}^{*}\mathbf{A}_{t}^{*}\mathbf{\bar{\Lambda }}(n)\text{+}\mathbf{B}_{r}^{*}{{\mathbf{B}}_{t}}\mathbf{\Lambda }(n)  \\
\end{matrix} \right].
\end{equation}

Hence, the $2\times2$ STBC transmitted data matrices $\mathbf{\hat{S}}_{k}(n)$
and $\mathbf{\hat{\bar{S}}}_{k}\text{(}n\text{)}$ corresponding to
the $n$-th and $(N+2-n)$-th subcarrier of the $\left(2k+1\right)$-th
and $\left(2k+2\right)$-th OFDM symbols, respectively, can be jointly
recovered as follows 
\begin{equation}
\left[\begin{matrix}\mathbf{\hat{S}}_{k}(n)\\
\mathbf{\hat{\bar{S}}}_{k}\text{(}n\text{)}
\end{matrix}\right]=\underbrace{\left[\begin{matrix}\mathbf{\Gamma}_{11}\text{(}n\text{)} & \mathbf{\Gamma}_{12}\text{(}n\text{)}\\
\mathbf{\Gamma}_{21}\text{(}n\text{)} & \mathbf{\Gamma}_{22}\text{(}n\text{)}
\end{matrix}\right]}_{\mathbf{\Gamma}\text{(}n\text{)}}\left[\begin{matrix}\mathbf{\mathbf{Z'}}_{k}\left(n\right)\\
\mathbf{\mathbf{\bar{Z}'}}_{k}\left(n\right)
\end{matrix}\right].
\end{equation}

In the absence of noise, the transmitted symbols can be perfectly
recovered when $\mathbf{\Gamma}(n)=\mathbf{\Phi}^{-1}(n)$. However,
since CSI is unknown in DSTBC-OFDM, it is not possible to invert $\mathbf{\Phi}(n)$.
Thus, a new strategy is needed to estimate and compensate IQI in DSTBC-OFDM.
Unlike coherent systems, we do not need to recover the exact transmitted
signal, instead, we only need to ensure that the differential encoding
relationship in Eq. (\ref{difsu-1}) is still satisfied by the adjacent
data blocks. However, by examining Eqs. (\ref{eq:freDiff_k_IQ}) and
(\ref{eq:eqfrediff1_k+1_IQ}), we find that the differential encoding
relationship no longer holds in the presence of IQI even without noise,
$i.e.$ 
\begin{equation}
\left[\begin{matrix}\mathbf{\mathbf{Z'}}_{k+1}\left(n\right)\\
\mathbf{\mathbf{\bar{Z}'}}_{k+1}\left(n\right)
\end{matrix}\right]\ne\left[\begin{matrix}\mathbf{\mathbf{Z'}}_{k}\left(n\right)\mathbf{U}_{k+1}\text{(}n\text{)}\\
\mathbf{\mathbf{\bar{Z}'}}_{k}\left(n\right)\mathbf{\bar{U}}_{k+1}\text{(}n\text{)}
\end{matrix}\right].
\end{equation}

It can be verified that the necessary condition to satisfy this relationship
is given by 
\begin{equation}
\mathbf{\Gamma}\text{(}n\text{)}\mathbf{\Phi}\text{(}n\text{)}=\left[\begin{matrix}\mathbf{H}_{1} & \mathbf{0}_{2\times2}\\
\mathbf{0}_{2\times2} & \mathbf{H}_{2}
\end{matrix}\right],\label{Dig}
\end{equation}
where $\mathbf{H}_{i}$ ($i=1,2$) are non-unique Alamouti matrices
that are related to the channel and IQI parameters. Hence, the following
relations must hold{\small{}
\begin{gather}
\left\{ \begin{aligned}\mathbf{\Gamma}_{11}(\mathbf{A}_{r}\mathbf{B}_{t}\mathbf{\Lambda}(n)\text{+}\mathbf{A}_{t}^{*}\mathbf{B}_{r}\mathbf{\bar{\Lambda}}(n))\thinspace\thinspace\thinspace\thinspace\thinspace\thinspace\thinspace\thinspace\thinspace\thinspace\thinspace\thinspace\thinspace\thinspace\thinspace\thinspace\\
+\mathbf{\Gamma}_{12}(\mathbf{A}_{r}^{*}\mathbf{A}_{t}^{*}\mathbf{\bar{\Lambda}}(n)\text{+}\mathbf{B}_{r}^{*}\mathbf{B}_{t}\mathbf{\Lambda}(n))=0,\\
\mathbf{\Gamma}_{21}(\mathbf{A}_{r}\mathbf{A}_{t}\mathbf{\Lambda}(n)\text{+}\mathbf{B}_{r}\mathbf{B}_{t}^{*}\mathbf{\bar{\Lambda}}(n))\thinspace\thinspace\thinspace\thinspace\thinspace\thinspace\thinspace\thinspace\thinspace\thinspace\thinspace\thinspace\thinspace\thinspace\thinspace\thinspace\\
+\mathbf{\Gamma}_{22}(\mathbf{A}_{r}^{*}\mathbf{B}_{t}^{*}\mathbf{\bar{\Lambda}}(n)\text{+}\mathbf{A}_{t}\mathbf{B}_{r}^{*}\mathbf{\Lambda}(n))=0.
\end{aligned}
\right.\label{qrs}
\end{gather}
}{\small \par}

Since any non-zero matrix $\mathbf{\Gamma}\text{(}n\text{)}$ which
satisfies the relations in Eq. (\ref{qrs}) satisfies the differential
encoding property, we set $\mathbf{\Gamma}_{11}=\mathbf{\Gamma}_{22}=\mathbf{I}$
for simplicity. Thus, we only need to satisfy the following condition
{\small{}
\begin{align}
\mathbf{\Gamma}_{12} & =\bar{\mathbf{\Gamma}}_{21}\triangleq\mathbf{\Gamma}_{c}(n)\nonumber \\
 & =-(\mathbf{A}_{r}\mathbf{B}_{t}\mathbf{\Lambda}(n)\text{+}\mathbf{A}_{t}^{*}\mathbf{B}_{r}\mathbf{\bar{\Lambda}}(n))(\mathbf{A}_{r}^{*}\mathbf{A}_{t}^{*}\mathbf{\bar{\Lambda}}(n)\text{+}\mathbf{B}_{r}^{*}\mathbf{B}_{t}\mathbf{\Lambda}(n))^{-1}.\label{gammac-a}
\end{align}
}{\small \par}

Eq. (\ref{gammac-a}) shows that $\mathbf{\Gamma}_{c}(n)$ varies
across the subcarriers. Hence, the recovered transmitted data matrices
$\mathbf{\hat{S}}_{k}(n)$ and $\mathbf{\hat{\bar{S}}}_{k}\text{(}n\text{)}$
can be re-formulated as follows
\begin{equation}
\left[\begin{matrix}\mathbf{\hat{S}}_{k}(n)\\
\mathbf{\hat{\bar{S}}}_{k}\text{(}n\text{)}
\end{matrix}\right]=\left[\begin{matrix}\mathbf{I} & \mathbf{\Gamma}_{c}\text{(}n\text{)}\\
\mathbf{\bar{\Gamma}}_{c}\text{(}n\text{)} & \mathbf{I}
\end{matrix}\right]\left[\begin{matrix}\mathbf{\mathbf{Z'}}_{k}\left(n\right)\\
\mathbf{\mathbf{\bar{Z}'}}_{k}\left(n\right)
\end{matrix}\right].\label{eq:compstructure}
\end{equation}

Similarly, the recovered $2\times2$ STBC transmitted data matrices
$\mathbf{\hat{S}}_{k+1}(n)$ and $\mathbf{\hat{\bar{S}}}_{k+1}\text{(}n\text{)}$
corresponding to the $\left(2\left(k+1\right)+1\right)$-th, $\left(2\left(k+1\right)+2\right)$-th
OFDM symbols using Eq. (\ref{WLTRIQ-1-1}) can be expressed as follows
\begin{equation}
\left[\begin{matrix}\mathbf{\hat{S}}_{k+1}(n)\\
\mathbf{\hat{\bar{S}}}_{k+1}\text{(}n\text{)}
\end{matrix}\right]=\left[\begin{matrix}\mathbf{I} & \mathbf{\Gamma}_{c}\text{(}n\text{)}\\
\mathbf{\bar{\Gamma}}_{c}\text{(}n\text{)} & \mathbf{I}
\end{matrix}\right]\left[\begin{matrix}\mathbf{\mathbf{Z'}}_{k+1}\left(n\right)\\
\mathbf{\mathbf{\bar{Z}'}}_{k+1}\left(n\right)
\end{matrix}\right].
\end{equation}

Since there is no training phase in differential transmission, the
estimation of the parameter ${{\mathbf{\Gamma}}_{c}}(n)$ to compensate
IQI at the $n$-th OFDM subcarrier can only be done based on the received
signal. We propose a decision-directed algorithm to estimate the compensation
parameter ${{\mathbf{\Gamma}}_{c}}(n)$. A least-mean-squares estimation
of the compensation matrix ${{\mathbf{\Gamma}}_{c}}(n)$ can be realized
as follows{\small{}
\begin{equation}
\begin{aligned}{{\mathbf{\Gamma}}_{c}}(n) & =\arg\ \underset{\mathbf{\Gamma}_{c}(n)}{\mathop{\min}}\,E\biggl\{\left|\mathbf{\mathbf{Z'}}_{k+1}(n)+{{{\mathbf{\Gamma}}_{c}}(n)}\mathbf{\mathbf{\bar{Z}'}}_{k+1}(n)\right.\\
 & \left.-\left(\mathbf{\mathbf{Z'}}_{k}(n)+{{\mathbf{\Gamma}}_{c}}(n)\mathbf{\mathbf{\bar{Z}'}}_{k}(n)\right)\mathbf{U}_{k+1}(n)\right|^{2}\biggr\}\\
 & =\arg\ \underset{{{\mathbf{\Gamma}}_{c}}(n)}{\mathop{\min}}\,E\Biggl\{\left|\underbrace{\mathbf{\mathbf{Z'}}_{k+1}(n)-\mathbf{\mathbf{Z'}}_{k}(n)\mathbf{U}_{k+1}(n)}_{\mathbf{\Xi}_{k}(n)}\right.\\
 & \left.+{{\mathbf{\Gamma}}_{c}}(n)\underbrace{(\mathbf{\mathbf{\bar{Z}'}}_{k+1}(n)-\mathbf{\mathbf{\bar{Z}'}}_{k}(n)\mathbf{U}_{k+1}(n))}_{\mathbf{\Delta}_{k}(n)}\right|^{2}\Biggr\}.
\end{aligned}
\label{eq:gammafull}
\end{equation}
}{\small \par}

The matrices $\mathbf{\Xi}_{k}(n)$ and $\mathbf{\Delta}_{k}(n)$
defined above have the orthogonal Alamouti structure. Thus, the estimation
can be simplified by considering only the $1^{st}$ columns of $\mathbf{\Xi}_{k}(n)$
and $\mathbf{\Delta}_{k}(n)$. Thus, Eq. (\ref{eq:gammafull}) can
be simplified as follows{\small{}
\begin{align}
{{\mathbf{\Gamma}}_{c}(n)} & =\arg\ \underset{{\mathbf{\mathbf{{\mathbf{\Gamma}}}}_{c}}}{\mathop{\min}}\,E\left\{ {\left|\left[\begin{matrix}[\mathbf{\Xi}_{k}(n)]{_{1,1}} & [\mathbf{\Xi}_{k}{(n)}]{_{2,1}}\end{matrix}\right]^{T}\right.}\right.\nonumber \\
 & \left.\left.+{\mathbf{\mathbf{{\mathbf{\Gamma}}}}}_{c}\left[\begin{matrix}{{[\mathbf{\Delta}_{k}(n)]}_{1,1}} & {{[\mathbf{\Delta}_{k}{(n)}]}_{2,1}}\end{matrix}\right]^{T}\right|^{2}\right\} ,\label{eq:gammasimple-1}
\end{align}
}where $\left[\begin{matrix}[\mathbf{\Xi}_{k}(n)]{_{1,1}} & [\mathbf{\Xi}_{k}{(n)}]{_{2,1}}\end{matrix}\right]^{T}$are
the elements of the $1^{st}$ column of the matrix $\mathbf{\Xi}_{k}(n)$
and $\left[\begin{matrix}{{[\mathbf{\Delta}_{k}(n)]}_{1,1}} & {{[\mathbf{\Delta}_{k}{(n)}]}_{2,1}}\end{matrix}\right]^{T}$
are the elements of the $1^{st}$ column of the matrix $\mathbf{\Delta}_{k}(n)$. 

Moreover, since ${{\mathbf{\Gamma}}_{c}}\text{(}n\text{)}$ is also
an Alamouti matrix as shown in Eq. \eqref{gammac-a}, we define the
elements in $\mathbf{\Gamma}_{c}(n)$ as 
\begin{equation}
\mathbf{\Gamma}_{c}\text{(}n\text{)}=\left[\begin{matrix}\gamma_{1}\text{(}n\text{)} & \gamma_{2}\text{(}n\text{)}\\
-\gamma_{2}\text{(}n\text{)}^{*} & \gamma_{1}\text{(}n\text{)}^{*}
\end{matrix}\right].
\end{equation}

After some simple manipulation, Eq. \eqref{eq:gammasimple-1} can
be further simplified to{\small{}
\begin{gather}
\begin{array}{c}
[\underbrace{\begin{array}{c}
\gamma_{1}(n)\thinspace\gamma_{2}(n)\end{array}]}\\
\mathbf{\boldsymbol{\gamma}}\text{(}n\text{)}
\end{array}=\arg\,\min\,E\left(\left|\underbrace{\left[\begin{array}{c}
\left[\mathbf{\Xi}(n)\right]_{1,1}\thinspace[\mathbf{\Xi}(n)^{*}]_{2,1}\end{array}\right]^{T}}_{\mathbf{\boldsymbol{\xi}}'(n)}\right.\right.\nonumber \\
\left.\left.+\boldsymbol{\mathbf{\gamma}}\text{(}n)\underset{\Delta'(n)}{\underbrace{\left[\begin{array}{cc}
{[\mathbf{\Delta}(n)]}_{1,1} & {[\mathbf{\Delta}(n)^{*}]}_{2,1}\\
{[\mathbf{\Delta}(n)]}_{2,1} & {-[\mathbf{\Delta}(n)^{*}]}_{1,1}
\end{array}\right]}}\right|^{2}\right).\label{52}
\end{gather}
}{\small \par}

We use the adaptive Recursive Least-Squares (RLS) algorithm to iteratively
estimate $\boldsymbol{\gamma}(n\text{)}$. We define $\boldsymbol{\gamma}_{m}(n)$
to be the estimated compensation vector for the $n$-th subcarrier
$\boldsymbol{\gamma}\text{(}n\text{)}$ after $m-1$ iterations. Then,
the $m$-th RLS iteration can be expressed as 
\begin{equation}
\begin{aligned}{{e}_{m}}(n) & \triangleq{{\boldsymbol{\mathbf{\xi}}}^{'}}(n)+{{{\mathbf{\boldsymbol{\gamma}}}'}_{m-1}}\text{(}n\text{)}{{\mathbf{\boldsymbol{\delta}}}^{'}}(n),\\
{{\mathbf{\boldsymbol{\gamma}}}_{m}(n)} & \triangleq{{\mathbf{\boldsymbol{\gamma}}}_{m-1}(n)}+\mathbf{k}_{m}^{H}e_{m},
\end{aligned}
\label{eq:RLS}
\end{equation}
where $\mathbf{k}_{m}=\mathbf{P}_{m-1}{{\mathbf{\boldsymbol{\delta}}}^{'}}(n)^{H}/\left(\mu+{{\mathbf{\boldsymbol{\delta}}}^{'}}(n){}^{H}\mathbf{P}_{m-1}{{\mathbf{\boldsymbol{\delta}}}^{'}}(n)\right)$
and $\mathbf{P}_{m}=\frac{1}{\mu}\left(\mathbf{P}_{m-1}-\mathbf{k}_{m}{{\boldsymbol{\mathbf{\delta}}}^{'}}(n)^{H}\mathbf{P}_{m-1}\right)$.
Also, the vector ${\mathbf{\boldsymbol{\xi}}}^{'}(n)$ is defined
in Eq. \eqref{52} and vector ${\mathbf{\boldsymbol{\delta}}}^{'}(n)$
is chosen from the $1^{st}$ and $2^{nd}$ columns of $\Delta'(n)$
defined in Eq. (\ref{52}). In addition, $\mu$ is the RLS adaptation
step size. 

\subsubsection{\label{subsec:RX-IQI-only}Special Case 1: Estimation and Compensation
for the RX-IQI-only Case}

In this subsection, we discuss a special case when there is no or
negligible TX-IQI, $i.e.$ ${{\mathbf{A}}_{t}}\approx\mathbf{I}$
and ${{\mathbf{B}}_{t}}\approx\mathbf{0}$ in Eq. \eqref{WLTRIQ-2}
and Eq. \eqref{WLTRIQ-1-1}. Consequently, the compensation matrix
${{\mathbf{\Gamma}}_{c}}$ in Eq. \eqref{gammac-a} reduces to the
diagonal matrix {\small{}
\begin{equation}
{{\mathbf{\Gamma}}_{12}}={{\mathbf{\Gamma}}_{21}^{*}}=-{\mathbf{B}_{r}}(\mathbf{A}_{r}^{*})^{-1}\triangleq\mathbf{\Gamma}_{r}=\left[\begin{array}{cc}
\gamma_{r} & 0\\
0 & \gamma_{r}^{*}
\end{array}\right],\label{eq:gamarr}
\end{equation}
}which is independent of the subcarrier index (and also the wireless
channel) and determined by a single scalar $\gamma_{r}$. Thus, all
subcarriers over all OFDM symbols have the same compensation matrix.
To estimate this matrix, we only need to estimate a scalar $\gamma_{r}=-{{\beta}_{r}}/{\alpha}_{r}^{*}$
and this is similar to \cite{7458826} and \cite{gc} which discuss
RX-IQI compensation. Consequently, by setting ${\boldsymbol{\mathbf{\gamma}}}=\left[\begin{array}{cc}
\gamma_{r} & 0\end{array}\right]$, after some manipulation, the estimator in Eq. \eqref{52} can be
simplified to {\small{}
\begin{eqnarray}
\gamma_{r} & = & \arg\ \underset{{\gamma_{r}}}{\mathop{\min}}\,E\left(\Biggl|\left[[\mathbf{\Xi}(n)]{}_{1,1}\thinspace\thinspace[\mathbf{\Xi}(n)^{*}]{}_{2,1}\right]\right.\nonumber \\
 &  & \left.+\gamma_{r}\left[\begin{matrix}[\mathbf{\Delta}(n)]_{1,1}\thinspace\thinspace[\mathbf{\Delta}(n)^{*}]_{2,1}\end{matrix}\right]\Biggr|^{2}\right),\label{56}
\end{eqnarray}
}which becomes a scalar estimation problem. On the other hand, since
the compensation parameter is the same for all the subcarriers, the
adaptive estimation of ${{\gamma}_{r}}$ should be done not only along
the time direction, but also in the frequency-domain along adjacent
subcarriers, thus the convergence speed is notably enhanced. 

To summarize, the RLS algorithm recursions to iteratively estimate
$\gamma_{r}$ are given by {\small{}
\begin{equation}
\begin{aligned}{{e}_{m}}(n) & \triangleq{\xi}(n)+{\gamma_{r,m-1}}{\mathbf{\delta}}(n),\\
{{\mathbf{\gamma}}_{r,m}} & \triangleq{{\mathbf{\gamma}}_{r,m-1}}+k_{m}^{*}e_{m},
\end{aligned}
\label{eq:RLS-1}
\end{equation}
}where the scalar $k_{m}$ is defined by $k_{m}=P_{m-1}{\mathbf{\delta}}(n)^{*}/\left(\mu+{\mathbf{\delta}}(n){}^{*}P_{m-1}{\mathbf{\delta}}(n)\right)$
and $P_{m}=\frac{1}{\mu}\left(P_{m-1}-k_{m}{\mathbf{\delta}}(n)^{*}P_{m-1}\right)$.
The scalar ${\mathbf{\gamma}}_{r,m-1}$ is the estimated compensation
parameter $\gamma_{r}$ after $m-1$ iterations and the set $(\mathbf{\xi}(n),\mathbf{\delta}(n))$
is chosen from the available set $\left\{ ([\mathbf{\Xi}_{k}(n)]{}_{1,1},[\mathbf{\Delta}_{k}(n)]_{1,1}),([\mathbf{\Xi}_{k}^{*}{(n)}]{}_{2,1},[\mathbf{\Delta}_{k}^{*}{(n)}]_{2,1})\right\} $
defined in Eq. (\ref{eq:gammafull}). 

\subsubsection{\label{subsec:SNR-degradation-TX-comp}Special Case 2: SNR Degradation
after Compensation for the TX-IQI-only case}

TX-IQI happens before noise and we are compensating IQI based on noisy
symbols. Hence, in the presence of TX-IQI (assume no RX-IQI), an inevitable
noise amplification will be introduced to the compensated symbol even
with perfect estimation of ${\mathbf{\Gamma}}_{c}\text{(}n)$, $i.e.$
${\mathbf{\Gamma}}_{c}\text{(}n)=(|\beta_{t}|\mathbf{\Lambda}(n))(|\alpha_{t}|\mathbf{\bar{\Lambda}}(n))^{-1}\triangleq{{\mathbf{\Gamma}}_{c}^{TX}}\text{(}n)$.
From Eqs. \eqref{gammac-a} and \eqref{eq:compstructure}, by replacing
the TX-IQI parameters with their absolute values, the received signal
after the TX-IQI compensation in the absence of noise is given by{\small{}
\begin{equation}
\begin{aligned}\hat{\mathbf{\mathbf{S}}}_{k,TX}\left(n\right) & =\mathbf{\mathbf{Z'}}_{k,TX}+\mathbf{\Gamma}_{c}^{TX}\mathbf{\mathbf{\bar{Z}'}}_{k,TX}\\
 & =|\alpha_{t}|\mathbf{\Lambda}(n)\mathbf{S}_{k}\left(n\right)+\mathbf{\Gamma}_{c}^{TX}\text{(}n\text{)}\left(|\beta_{t}|\bar{\mathbf{\Lambda}}(n)\right)\mathbf{S}_{k}\left(n\right)\\
 & =|\alpha_{t}|\left(1-\rho_{t}\right)\mathbf{\Lambda}(n)\mathbf{S}_{k}\left(n\right),
\end{aligned}
\label{eq:SNRtx}
\end{equation}
}which shows that the signal power is reduced by a factor of $|\alpha_{t}|^{2}\left(1-\rho_{t}\right)^{2}$.
On the other hand, the noise term in Eq. (\ref{WLTRIQ-2}) after perfect
compensation becomes{\small{}
\begin{eqnarray}
\mathbf{V}_{k,TX} & = & \mathbf{V}_{k}+\mathbf{\Gamma}_{c}^{TX}\text{(}n\text{)}\mathbf{\bar{V}}_{k}\nonumber \\
 & = & \mathbf{V}_{k}-(|\beta_{t}|\mathbf{\Lambda})(|\alpha_{t}|\mathbf{\bar{\Lambda}})^{-1}\mathbf{\bar{V}}_{k}.
\end{eqnarray}
}{\small \par}

Since the noise samples in the mirror subcarriers $\mathbf{V}_{k,TX}$
and $\mathbf{\bar{V}}_{k,TX}$ are independent zero-mean Gaussian
distributed, the noise power after compensation is{\small{}
\begin{align}
E\{Tr(\mathbf{V}_{k,TX}\mathbf{V}_{k,TX}^{H})/4\} & =E\{Tr(\mathbf{V}_{k}\mathbf{V}_{k}^{H})/4\nonumber \\
 & +E\{Tr\left((\mathbf{\mathbf{\bar{V}}}_{k}\mathbf{\Gamma}_{c}^{TX})(\mathbf{\mathbf{\bar{V}}}_{k}\mathbf{\Gamma}_{c}^{TX})^{H}\right)/4\nonumber \\
 & =(1+2\rho_{t})\sigma^{2},\label{eq:nosieamp}
\end{align}
}where we use the fact given before Eq. \eqref{eeta} that the ratio
of the channel gains $E\left(\mathbf{\Gamma}_{c}^{TX}\left(\mathbf{\Gamma}_{c}^{TX}\right)^{H}\right)=\rho_{t}E\left(\frac{\left|\lambda\right|^{2}}{|\overline{\lambda}|^{2}}\right)=2\rho_{t}$.
Eq. \eqref{eq:nosieamp} shows that even with perfect compensation,
the noise power will be amplified by the factor $(1+2\rho_{t})$.
Thus, the combined effect of Eq. \eqref{eq:nosieamp} and Eq. \eqref{eq:SNRtx}
results in a SNR degradation factor of $\frac{|\alpha_{t}|^{2}(1-\rho_{t})^{2}}{1+2\rho_{t}}\approx\frac{(1-\rho_{t})^{2}}{1+2\rho_{t}}<1$
even when a perfect compensation matrix is used.

\subsection{Parameter-based (PB) Estimation in the presence of TX-IQI under High-mobility\label{subsec:Estimation-of-TX-IQI}}

In Subsection \ref{subsec:Widely-linear-Compensation}, the compensation
matrix for each OFDM subcarrirer is estimated independently. In the
presence of TX-IQI, the compensation matrix is determined by both
the TX-IQI parameters and the CIR. However, in high-mobility scenarios,
the compensation performance will be significantly degraded. This
degradation is mainly caused by two factors: the first factor is the
ICI introduced by the Doppler effect. Both an IQI-free system and
an IQI-compensated system suffer from this degradation. The second
factor is that the compensation matrix changes with the time-varying
channel thus the adaptive estimation works on a non-stationary basis,
where an extra ``lag error'' is introduced \cite{RLS}. We concentrate
on the latter factor because it creates an SNR degradation gap between
the performance of the IQI-compensated system and the IQI-free system.
Note that in the presence of TX-IQI and high mobility, it is not possible
to totally eliminate the lag error brought by the non-stationarity.
However, an estimation with faster convergence rate can reduce the
degradation caused by the lag error because it suffers less from the
accumulated non-stationarity errors during its estimation process.
Hence, if we could enhance the convergence rate of the adaptive estimation
algorithm, better performance could be obtained in a fast fading scenario.
A straightforward approach to improve the convergence rate is to reduce
the forgetting factor of RLS estimation but at the price of robustness
against noise.

In this subsection, we present an extension of the compensation algorithm
in Subsection \ref{subsec:Widely-linear-Compensation} that enhances
the convergence speed by jointly estimating the compensation matrix
of a subcarrier and its image subcarrier with the help of the IQI
parameters. The joint estimation improves the convergence speed because
we now have only one compensation matrix to estimate over two subcarriers
thus the convergence speed is doubled. Additionally, we will also
show how to estimate the TX-IQI and RX-IQI parameters needed in the
estimation. 

\subsubsection{Connection between Compensation Matrices of Image Subcarriers}

According to Eq.\eqref{gammac-a}, the compensation matrix for the
$n$-th subcarrier is given by{\small{}
\begin{align}
\mathbf{\Gamma}_{c}(n) & =-(\mathbf{A}_{r}\mathbf{B}_{t}\mathbf{\Lambda}(n)\text{+}\mathbf{A}_{t}^{*}\mathbf{B}_{r}\mathbf{\bar{\Lambda}}(n))(\mathbf{A}_{r}^{*}\mathbf{A}_{t}^{*}\mathbf{\bar{\Lambda}}(n)\text{+}\mathbf{B}_{r}^{*}\mathbf{B}_{t}\mathbf{\Lambda}(n))^{-1}\nonumber \\
 & \approx-(\mathbf{A}_{r}\mathbf{B}_{t}\mathbf{\Lambda}(n)\text{+}\mathbf{A}_{t}^{*}\mathbf{B}_{r}\mathbf{\bar{\Lambda}}(n))(\mathbf{A}_{r}^{*}\mathbf{A}_{t}^{*}\mathbf{\bar{\Lambda}}(n))^{-1}\nonumber \\
 & =-\mathbf{A}_{r}\mathbf{B}_{t}\mathbf{\Lambda}(n)\mathbf{\bar{\Lambda}}(n)^{-1}\left(\mathbf{A}_{r}^{*}\mathbf{A}_{t}^{*}\right)^{-1}+\mathbf{B}_{r}\left(\mathbf{A}_{r}^{*}\right)^{-1}.\label{eq:taoc_approx}
\end{align}
}{\small \par}

Similarly, the compensation matrix for its image subcarrier, denoted
as $\mathbf{\bar{\Gamma}}_{c}(n)$ is given by{\small{}
\begin{align}
\mathbf{\bar{\Gamma}}_{c}(n) & \approx-\mathbf{A}_{r}\mathbf{B}_{t}\mathbf{\bar{\Lambda}}(n)\mathbf{\Lambda}(n)^{-1}\left(\mathbf{A}_{r}^{*}\mathbf{A}_{t}^{*}\right)^{-1}+\mathbf{B}_{r}\left(\mathbf{A}_{r}^{*}\right)^{-1}.\label{eq:taocb_approx}
\end{align}
}{\small \par}

The channel components $\mathbf{\Lambda}(n)\mathbf{\bar{\Lambda}}(n)^{-1}$
and $\mathbf{\bar{\Lambda}}(n)\mathbf{\Lambda}(n)^{-1}$ in Eqs. \eqref{eq:taoc_approx}
and \eqref{eq:taocb_approx} are mutually conjugate inverse matrices,
$i.e.$ {\small{}$\left[\mathbf{\Lambda}(n)\left(\mathbf{\bar{\Lambda}}(n)^{-1}\right)\right]^{*}\mathbf{\bar{\Lambda}}(n)\mathbf{\Lambda}(n)^{-1}=\mathbf{I}$}
. Thus, the following constraint relationship between the two compensation
matrices can be derived{\small{}
\begin{equation}
\mathbf{\bar{\Gamma}}_{c}(n)-\mathbf{\Gamma}_{r}=\mathbf{\Gamma}_{t}\left[\left(\mathbf{\Gamma}_{c}(n)-\mathbf{\Gamma}_{r}\right)^{*}\right]^{-1}\mathbf{\Gamma}_{t}^{*},\label{eq:constraint}
\end{equation}
}where $\mathbf{\Gamma}_{t}=diag(\gamma_{t},\gamma_{t}^{*})$, $\mathbf{\Gamma}_{r}=diag(\gamma_{r},\gamma_{r}^{*})$,
$\gamma_{t}={{\beta}_{t}}/{\alpha}_{t}$ and $\gamma_{r}=\beta_{r}/\alpha_{r}^{*}$.

Eq. \eqref{eq:constraint} gives us a constraint between $\mathbf{\Gamma}_{c}(n)$
and $\mathbf{\bar{\Gamma}}_{c}(n)$. It only contains the compensation
matrices and the IQI parameters and is independent of the channel
realization. Hence, if we know the IQI parameters, the compensation
matrix of a given subcarrier can be obtained by the compensation matrix
of its image subcarrier.

\subsubsection{Estimation of the TX-IQI and RX-IQI Parameters}

In the remainder of this section, we assume that the compensation
matrices of Subcarrier $n$ and its image subcarrier, $i.e.$ $\mathbf{\Gamma}_{c}(n)$
and $\mathbf{\bar{\Gamma}}_{c}(n)$, are already estimated by the
algorithm described in Subsection \ref{subsec:Widely-linear-Compensation}.
Theoretically, when $\mathbf{\Gamma}_{c}(n)$ and $\mathbf{\bar{\Gamma}}_{c}(n)$
are already known, according to Eq. (\ref{eq:constraint}), the IQI
parameters could be straightforwardly estimated by solving the following
optimization problem{\small{}
\begin{equation}
[\gamma_{t},\gamma_{r}]=\underset{\gamma_{t},\gamma_{r}}{\text{arg}\thinspace\text{min}}\left|\mathbf{\bar{\Gamma}}_{c}(n)-\mathbf{\Gamma}_{r}-\mathbf{\Gamma}_{t}\left[\left(\mathbf{\Gamma}_{c}(n)-\mathbf{\Gamma}_{r}\right)^{*}\right]^{-1}\mathbf{\Gamma}_{t}^{*}\right|^{2}.\label{parTR}
\end{equation}
}{\small \par}

Note that the TX-IQI and RX-IQI parameters are often treated as time-invariant
parameters because they change very slowly with time. Hence, the convergence
speed of the IQI-parameter estimation process is not an important
issue any more. However, since Eq. \eqref{parTR} involves inverting
a matrix, although it enjoys an Alamouti structure, it is still very
complicated to solve the optimization problem directly. As a result,
we estimate the RX-IQI parameter $\gamma_{r}$ in an alternative way
and then simplify the cost function before estimating the TX-IQI parameter
$\gamma_{t}$ . Recall that $\mathbf{\Gamma}_{c}(n)$ could be written
as{\small{}
\begin{equation}
\mathbf{\Gamma}_{c}(n)=\mathbf{\Gamma}_{r}-\underset{n_{TX}}{\underbrace{\mathbf{A}_{r}\mathbf{B}_{t}\mathbf{\Lambda}(n)\mathbf{\bar{\Lambda}}(n)^{-1}\left(\mathbf{A}_{r}^{*}\mathbf{A}_{t}^{*}\right)^{-1}}},
\end{equation}
}where the term {\small{}$n_{TX}=-\mathbf{A}_{r}\mathbf{B}_{t}\mathbf{\Lambda}(n)\mathbf{\bar{\Lambda}}(n)^{-1}\left(\mathbf{A}_{r}^{*}\mathbf{A}_{t}^{*}\right)^{-1}$},
which is introduced by TX-IQI, can be treated as zero-mean noise.
Thus, $\mathbf{\Gamma}_{c}(n)$ can be regarded as a noisy version
of $\mathbf{\Gamma}_{r}$. However, since the power of term $n_{TX}$
could be equally strong or even stronger than $\mathbf{\Gamma}_{r}$,
direct estimation of $\mathbf{\Gamma}_{r}$ based on $\mathbf{\Gamma}_{c}(n)$
can be of poor accuracy even when enough samples of $\mathbf{\Gamma}_{c}(n)$
are collected. Fortunately, the power of interference term $n_{TX}$,
denoted as $E_{nTX}$, can be well predicted and it allows us to use
a weighted estimator to improve accuracy.

Define $\kappa(n)$ as the ratio of the received signal matrix power
in Subcarrier $n$ and that of its image subcarrier, $i.e.$ {\small{}
\begin{equation}
\kappa(n)=\mathbf{|\mathbf{Z'}}_{k}\left(n\right)|^{2}/|\mathbf{\mathbf{\bar{Z}'}}_{k}\left(n\right)|^{2}.
\end{equation}
}Since the IQI terms are generally small and the transmitted signal
matrix is normalized, we have {\small{}
\begin{equation}
\kappa(n)\approx|\mathbf{\lambda}(n)/\bar{\mathbf{\lambda}}(n)|^{2}\varpropto E_{nTX}=|\mathbf{\lambda}(n)/\bar{\mathbf{\lambda}}(n)|^{2}\rho_{t}.\label{eq:ratioT}
\end{equation}
}Thus, for a given subcarrier, a smaller $\kappa(n)$ indicates that
the RX-IQI compensation parameters are more dominant in $\mathbf{\Gamma}_{c}(n)$
and result in a more reliable observation of $\mathbf{\Gamma}_{r}$
(or $\gamma_{r}$). Hence, after simplifications based on the diagonal
Alamouti structure of $\mathbf{\Gamma}_{c}(n)$, $\gamma_{r}$ can
be estimated by a weighted adaptive estimation as follows{\small{}
\begin{equation}
\gamma_{r}=\gamma_{r}+\mu_{r}/\kappa(n)(\left[\mathbf{\Gamma}_{c}(n)\right]_{1,1}-\gamma_{r})\label{parR}
\end{equation}
}where $\mu_{r}$ is the step size for RX-IQI parameter estimation.

After $\gamma_{r}$ is obtained, the matrices $\mathbf{\Gamma}_{c}(n)-\mathbf{\Gamma}_{r}$
and $\mathbf{\bar{\Gamma}}_{c}(n)-\mathbf{\Gamma}_{r}$ become known
matrices. Define $\mathbf{\boldsymbol{\omega}}_{1}=[\mathbf{\omega}_{11,}\mathbf{\omega}_{12}]$
and $\mathbf{\boldsymbol{\omega}}_{2}=[\mathbf{\omega}_{21,}\mathbf{\omega}_{22}]$,
which are the first rows of {\small{}$\mathbf{\Gamma}_{c}(n)-\mathbf{\Gamma}_{r}$}and
{\small{}$\left[(\mathbf{\bar{\Gamma}}_{c}(n)-\mathbf{\Gamma}_{r})^{-1}\right]^{*}$},
respectively. Since {\small{}$\mathbf{\Gamma}_{c}(n)-\mathbf{\Gamma}_{r}$}
and {\small{}$\left[(\mathbf{\bar{\Gamma}}_{c}(n)-\mathbf{\Gamma}_{r})^{-1}\right]^{*}$}
are Alamouti matrices and $\mathbf{\Gamma}_{t}$ is a diagonal Alamouti
matrix, Eq. \eqref{parTR} can be simplified to
\begin{align}
\gamma_{t} & =\underset{\gamma_{t}}{\text{arg}\thinspace\text{min}}\left|[\mathbf{\omega}_{11\thinspace,\thinspace}\mathbf{\omega}_{12}]-[|\gamma_{t}|^{2}\mathbf{\omega}_{11}\thinspace,\thinspace\gamma_{t}{}^{2}\mathbf{\omega}_{12}]\right|^{2}.\label{parT-1}
\end{align}

The optimization problem in Eq. \eqref{parT-1} can be adaptively
solved by a simple recursive estimation of $A_{\gamma2}^{(n)}=|\gamma_{t}|^{2}$
and $\phi_{\gamma2}^{(n)}$=$\gamma_{t}{}^{2}/|\gamma_{t}|^{2}$ as
follows{\small{}
\begin{align}
A_{\gamma2}^{'}=A_{\gamma2}^{(n-1)}-\mu_{a}\nabla_{1}, & \thinspace\thinspace A_{\gamma2}^{(n)}=A_{\gamma2}^{'}-\mu_{a}\nabla_{2},\nonumber \\
\phi_{\gamma2}^{'}=\phi_{\gamma2}^{(n-1)}-\mu_{b}\nabla_{3}, & \thinspace\thinspace\phi_{\gamma2}^{(n)}=\text{\ensuremath{\phi_{\gamma2}^{'}}}/|\phi_{\gamma2}^{'}|,
\end{align}
}where {\small{}$\nabla_{1}=2A_{\gamma2}^{(n-1)}|\omega_{12}|^{2}-2\text{Re}\{\omega_{11}^{*}\omega_{12}\}$},
{\small{}$\nabla_{2}=2A_{\gamma2}^{'}|\omega_{22}|^{2}-2\text{Re}\{\omega_{21}^{*}\omega_{22}\phi_{\gamma2}^{(n-1)}\}$},
{\small{}$\nabla_{3}=-2\omega_{12}\omega_{22}^{*}A_{\gamma2}^{(n)}(\phi_{\gamma2}^{(n-1)})^{*}$},
$\mu_{a}$ and $\mu_{b}$ are the step sizes for the estimation of
the modulus and phase of the TX-IQI parameter, respectively.

Note that, although it is beyond the scope of this paper, in some
applications, there is a feedback link from the receiver to the transmitter.
In this case, an easy but effective way to compensate the TX-IQI is
to estimate TX-IQI parameter $\gamma_{t}$ at the receiver and then
send it back to the transmitter. After that, the TX-IQI could be eliminated
by applying a simple linear pre-distortion compensation at the transmitter,
where $b(n)$ is pre-distorted to $b(n)-\gamma_{t}b(n)^{*}$ before
transmitted. 

\subsubsection{PB estimation }

After the TX-IQI and RX-IQI parameters are estimated, Eq. \eqref{eq:constraint}
can be used to jointly estimate the compensation matrices $\mathbf{\Gamma}_{c}(n)$
and $\mathbf{\bar{\Gamma}}_{c}(n)$. To simplify our notation, we
assume that the RX-IQI is already compensated with the estimated parameter
$\gamma_{r}$ by applying it to the compensation matrix shown in Eq.
\eqref{eq:gamarr}. As a result, according to Eq. \eqref{eq:constraint},
the TX-IQI compensation matrices should satisfy{\small{}
\begin{align}
\mathbf{\bar{\Gamma}}_{c}(n) & =\mathbf{\Gamma}_{t}\left[\left(\mathbf{\Gamma}_{c}(n)\right)^{*}\right]^{-1}\mathbf{\Gamma}_{t}^{*},
\end{align}
}which is equivalent to {\small{}
\begin{equation}
\begin{cases}
\left[\mathbf{\bar{\Gamma}}_{c}(n)\right]_{1,1}=|\gamma_{t}|^{2}\left[\left(\mathbf{\Gamma}_{c}^{*}(n)\right)^{-1}\right]_{1,1},\\
\left[\mathbf{\bar{\Gamma}}_{c}(n)\right]_{1,2}=\gamma_{t}{}^{2}\left[\left(\mathbf{\Gamma}_{c}^{*}(n)\right)^{-1}\right]_{1,2}\thinspace.
\end{cases}\label{eq:n2nbr}
\end{equation}
}{\small \par}

Note that $\mathbf{\Gamma}_{c}^{*}(n)$ is a 2-by-2 Alamouti matrix;
therefore, it is very simple to calculate its inverse matrix. To estimate
$\mathbf{\Gamma}_{c}$ for a given pair of Subcarriers $n$ and $N-n+2$,
we first choose one of the two subcarriers and denote it by Subcarrier
$n^{\dagger}$. To obtain better performance, the choice of $n^{\dagger}$
should not be arbitrary. If we start by estimating $\mathbf{\Gamma}_{c}(n^{\dagger})$,
the estimation process will be ended by calculating the compensation
matrix $\mathbf{\Gamma}_{c}(n^{\dagger})$ with the inverse of $\mathbf{\Gamma}_{c}(N-n^{\dagger}+2)$,
which implies that a larger $\mathbf{\Gamma}_{c}(N-n^{\dagger}+2)$
is more robust to the noise and less error will be introduced by its
inverse. Consequently, we choose $n^{\dagger}$ in order to have a
larger power of $\mathbf{\Gamma}_{c}(N-n^{\dagger}+2)$. According
to Eq. \eqref{eq:ratioT}, the power $\mathbf{\Gamma}_{c}(n)$ can
be predicted by $\kappa(n)$. Hence $n^{\dagger}$ can be chosen as{\small{}
\begin{equation}
n^{\dagger}=\begin{cases}
n & \kappa(n)\leq1,\\
N-n+2 & \text{otherwise}.
\end{cases}\label{eq:deciden+}
\end{equation}
}{\small \par}

After $n^{\dagger}$ is determined, $\mathbf{\Gamma}_{c}(n^{\dagger})$
is estimated by Eq. \eqref{eq:RLS} and the symbols in Subcarriers
$n^{\dagger}$ and $\mathbf{\Gamma}_{c}(N-n^{\dagger}+2)$ are then
calculated with $\mathbf{\Gamma}_{c}(n^{\dagger})$ using Eq. \eqref{eq:n2nbr}.
After that, $\mathbf{\Gamma}_{c}(N-n^{\dagger}+2)$ is updated again
using Eq. \eqref{eq:RLS} with the data in Subcarrier $N-n^{\dagger}+2$.
Finally, $\mathbf{\Gamma}_{c}(n^{\dagger})$ is calculated using Eq.
\eqref{eq:n2nbr} with the updated $\mathbf{\Gamma}_{c}(N-n^{\dagger}+2)$.
Therefore, the compensation matrices $\mathbf{\Gamma}_{c}(n^{\dagger})$
and $\mathbf{\Gamma}_{c}(N-n^{\dagger}+2)$ are updated twice within
one DSTBC-OFDM block and the convergence speed is thus doubled. In
addition, since the PB estimation process requires knowledge of the
IQI parameters which must be estimated using the WL estimation, it
may seem that two different algorithms must be implemented in the
receiver. However, the estimation mechanisms of the two algorithms
are highly overlapped. In fact, the major difference between them
lies only in the estimation of the IQI parameters, which is basically
a one-time operation because the IQI parameters are almost time-invariant
and once they are estimated, they would be valid for a long period
of time. 

Assume that the first $N_{ini}$ received symbols are used for estimating
the IQI parameters, the PB estimation and compensation algorithm is
summarized in Algorithm \ref{enu:a1}.

\begin{algorithm}[tb]
\textbf{Input:} $\mathbf{\mathbf{Z'}}\left(n\right)$, $\mathbf{\mathbf{\bar{Z}'}}\left(n\right)$,
$t=0$

\textbf{Output}: $\mathbf{\hat{\bar{S}}}\text{(}n\text{)}$, $\mathbf{\hat{\bar{S}}}\text{(}n\text{)}$\textbf{\vspace{-0.5cm}}
\begin{enumerate}
\begin{singlespace}
\item \textbf{if} $t\leq N_{ini}$
\end{singlespace}
\item \quad{}update {\small{}$\mathbf{\Gamma}_{c}(n)$} and {\small{}$\bar{\mathbf{\Gamma}}_{c}(n)$}
using Eq. \eqref{eq:RLS};
\item \quad{}update $\gamma_{t}$ and $\gamma_{r}$ using Eqs. \eqref{parR}
and \eqref{parT-1};
\item \textbf{else}
\item \quad{}determine $n^{\dagger}$ using Eq. \eqref{eq:deciden+};
\item \quad{}update {\small{}$\mathbf{\Gamma}_{c}(n^{\dagger})$} using
Eq. \eqref{eq:RLS};
\item \quad{}calculate $\mathbf{\Gamma}_{c}(N-n^{\dagger}+2)$ using Eq.
\eqref{eq:n2nbr};
\item \quad{}update {\small{}$\mathbf{\Gamma}_{c}(N-n^{\dagger}+2)$} using
Eq. \eqref{eq:RLS};
\item \quad{}calculate $\mathbf{\Gamma}_{c}(n^{\dagger})$ using Eq. \eqref{eq:n2nbr};
\item \textbf{end if}
\item compensate IQI using Eq. \eqref{eq:compstructure};
\item $t=t+1$, move on to next subcarrier;
\end{enumerate}
\caption{the PB estimation and compensation}
\label{enu:a1}
\end{algorithm}

\section{Numerical Results}

\label{sec:numerical} The system parameters are almost similar to
\cite{narasimhan2010reduced}. The transmitter sends QPSK and 8-PSK
modulated symbols over a bandwidth of 5MHz and the carrier frequency
is 2.5GHz. The number of OFDM subcarrier is set to 64. The slow-fading
channel model used is a Rayleigh fading channel with equal-power CIR
taps. We also examined the performance of the proposed blind IQI compensation
algorithm in a fast-fading channel, where the ITU Vehicular channel
A (ITU-VA) model is adopted. The mobile speed is 200km/h for fast
fading, corresponding to a maximum Doppler shift of 463Hz. Two levels
of TX-IQI and RX-IQI are considered in our simulation, which are moderate
IQI with amplitude imbalance ${{\kappa}_{t/r}}(dB)=0.5\mathrm{dB}$,
phase imbalance ${{\phi}_{t/r}}={{3}^{\circ}}$, and severe IQI with
${{\kappa}_{t/r}}=1\mathrm{dB}$, ${{\phi}_{t/r}}={{5}^{\circ}}$,
resulting in a transmitter/receiver IRR of 11.6dB and 18dB, respectively.
For both the WL estimation and PB estimation algorithms, the forgetting
factor of RLS algorithm $\mu$ is set to 0.9. In the PB estimation,
the first 1300 DSTBC-OFDM symbols (650 DSTBC-OFDM Alamouti codewords)
are used to estimate the TX-IQI parameter $\gamma_{t}$ and RX-IQI
parameter $\gamma_{r}$. The step size for estimation of the amplitude
of TX-IQI parameter, the phase of the TX-IQI parameter and the RX-IQI
parameter are set to $\mu_{a}=0.0005$, $\mu_{b}=0.005$, $\mu_{r}=0.0001$,
respectively.

\begin{figure}[tb]
\centering\includegraphics[width=3.5in]{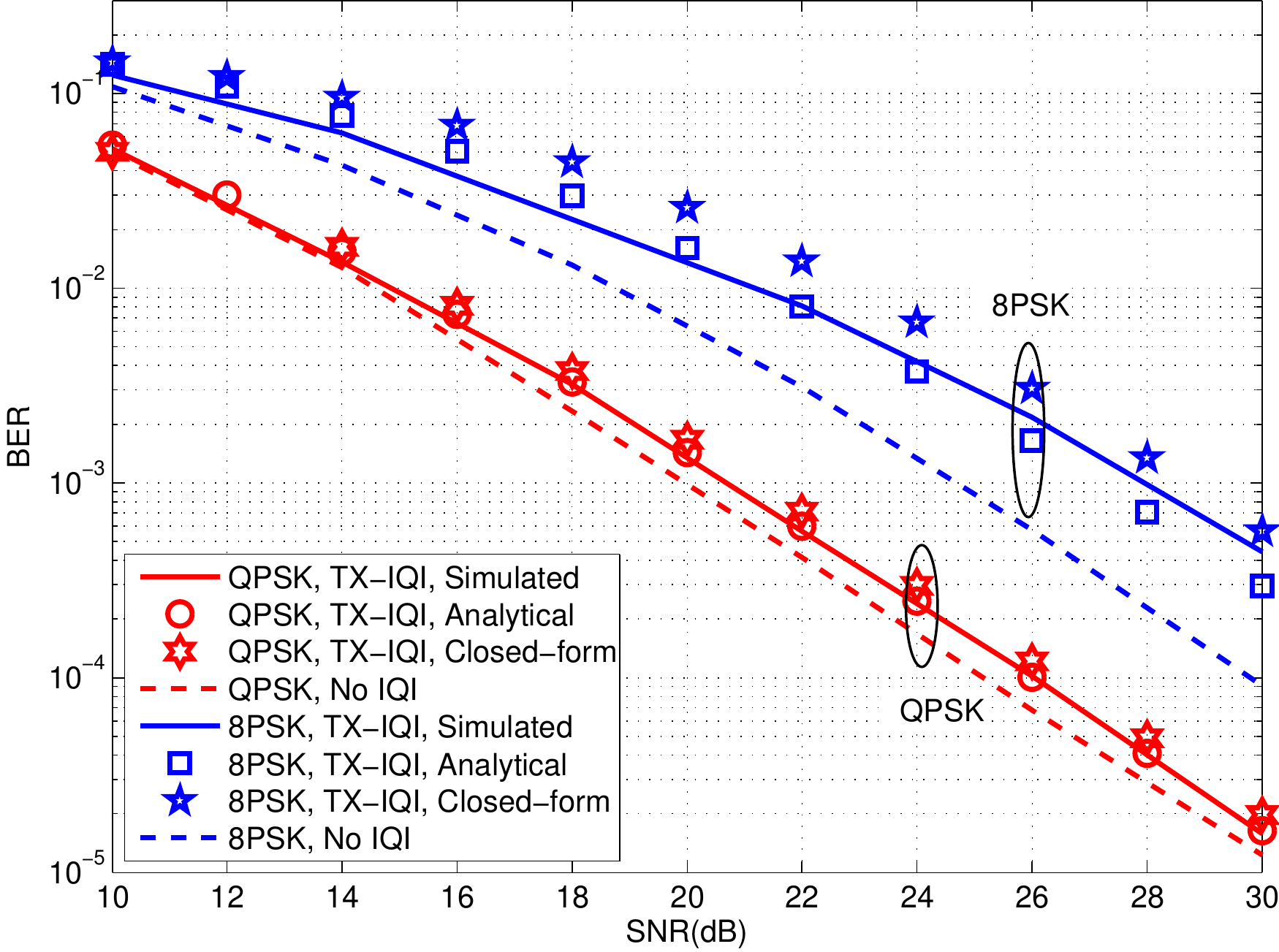}\vspace{-0.5cm}\caption{Comparison of analytical and simulated BER performance of the DSTBC-OFDM
system under TX-IQI ({\small{}slow fading, ${{\kappa}_{t}}(dB)=0.5\mathrm{dB}$
and ${{\phi}_{t}}={{3}^{\circ}}$}). \vspace{-0.2cm}}
\label{fig:analyticalber-t}
\end{figure}

Fig. \ref{fig:analyticalber-t} shows the analytical and closed-form
BER of QPSK and 8PSK in the presence of TX-IQI in Eq. \eqref{eq:BER-TX-analytical}
and Eq. \eqref{eq:BER-TX-closedform}, and compares them with the
simulated BER. The TX-IQI in simulation is set as the moderate case
with ${{\kappa}_{t}}(dB)=0.5\mathrm{dB}$, ${{\phi}_{t}}={{3}^{\circ}}$.
The analytical BER of QPSK matches the simulated BER while a small
gap is observed in the 8PSK case. The gap is due to both the inaccuracy
in modeling severe TX-IQI for the high SINR case, and the Taylor expansion
approximation in Eq. \eqref{eq:amploss} because the TX-IQI we assume
in the simulation is severe for a first-order Taylor approximation.
In addition, according to Eq. \eqref{eq:gap}, the SNR loss caused
by TX-IQI should be $0.70$ dB for QPSK and $2.68$ dB for 8PSK, which
also matches the simulation results and confirms that 8PSK is less
robust to TX-IQI than QPSK, as expected.

\begin{figure}[tb]
\centering\includegraphics[width=3.5in]{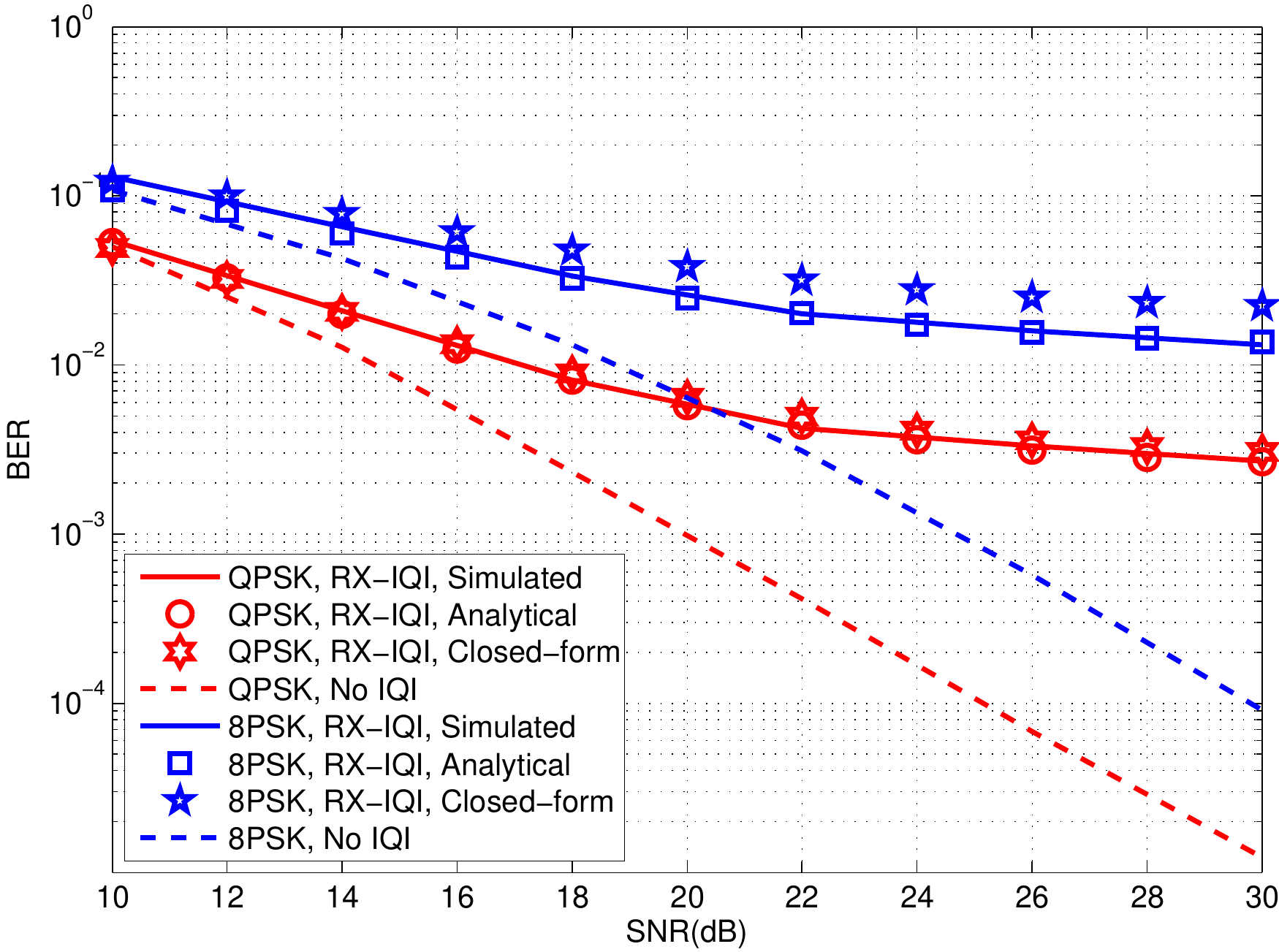}\vspace{-0.5cm}\caption{Comparison of analytical and simulated BER performance of the DSTBC-OFDM
system under RX-IQI ({\small{}slow fading, ${{\kappa}_{t}}(dB)=0.5\mathrm{dB}$,
${{\phi}_{t}}={{3}^{\circ}}$}).\vspace{-0.5cm} }
\label{fig:analyticalber-r}
\end{figure}

\begin{figure}[tb]
\centering\includegraphics[width=3.5in]{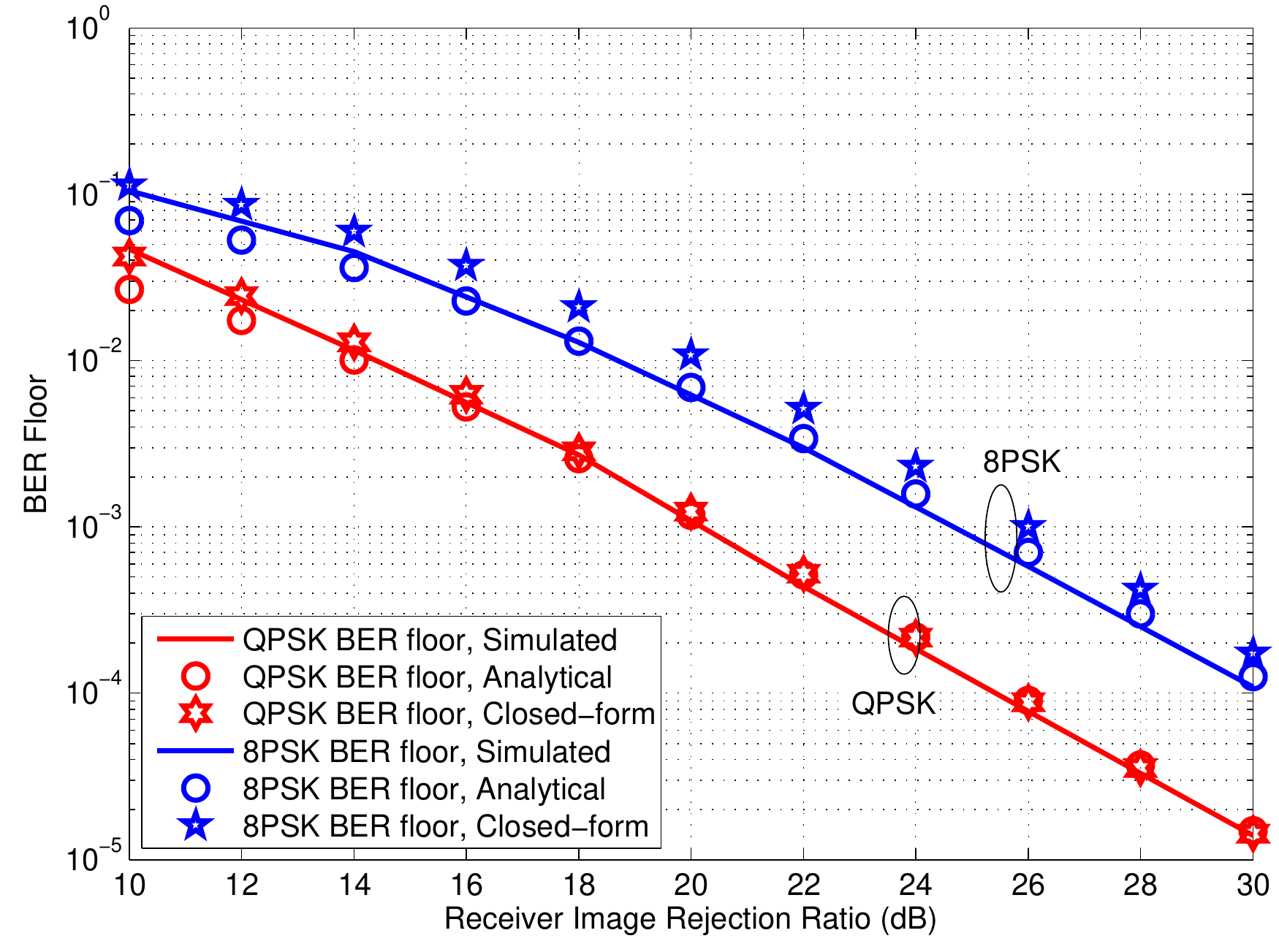}\vspace{-0.5cm}\caption{Comparison of analytical and simulated BER floor of the DSTBC-OFDM
system under RX-IQI ({\small{}slow fading, ${{\kappa}_{r}}(dB)=0.5\mathrm{dB}$
and ${{\phi}_{r}}={{3}^{\circ}}$}).\vspace{-0.5cm}}
\label{fig:IRR ber}
\end{figure}

Fig. \ref{fig:analyticalber-r} compares the analytical BER in Eq.
\eqref{BER-RX-analytical} with the SINR distribution in Eq. \eqref{eq:pEtaRX}
and the closed-form BER in Eq. \eqref{eq:BER-RX-closedform}. Both
of them match the simulated BER results for the RX-IQI case. The RX-IQI
parameters are set as the moderate case. It is clear that the BER
is much more sensitive to RX-IQI than TX-IQI as shown in Fig. \ref{fig:analyticalber-t}.
The BER curves of both QPSK and 8PSK show a BER floor in the high
SNR region, which is caused by the limited SINR even in the absence
of noise. Fig. \ref{fig:IRR ber} shows the BER floor of QPSK and
8PSK under RX-IQI for different IRR scenarios. The analytical BER
is obtained from Eq. \eqref{BER-RX-analytical} and the closed-form
BER floor is obtained from Eq. \eqref{eq:BER-RX-closedform}. As shown
in Subsection \ref{subsec:DSTBC-OFDM-under-RX-IQI}, the BER floors
of QPSK and 8PSK for different IRR levels show a $5.5$dB gap in the
high IRR region.

\begin{figure}[tb]
\centering\includegraphics[width=3.5in]{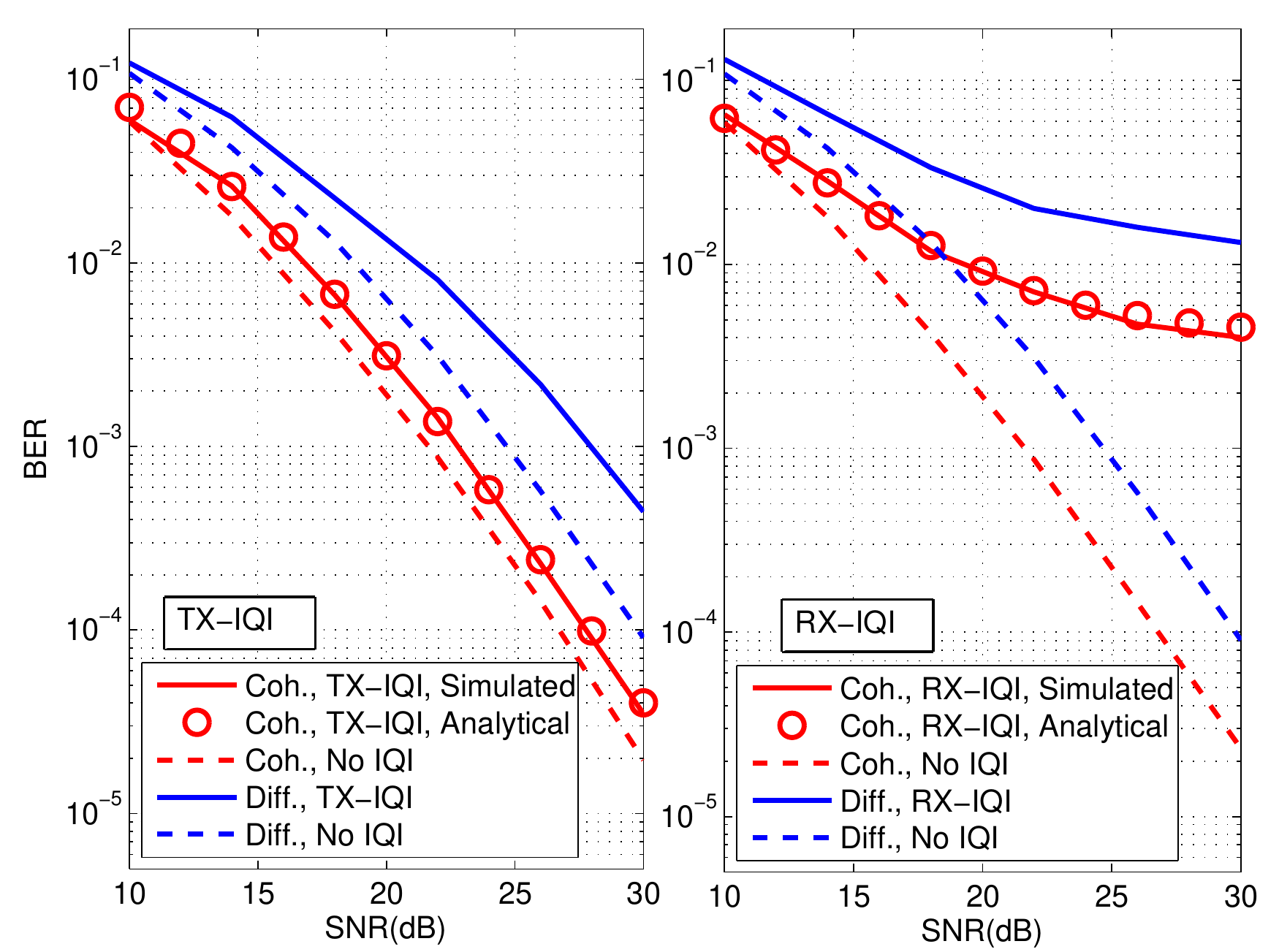}\vspace{-0.5cm}\caption{Comparison of BER performance of the differential system and coherent
system under IQI ({\small{}slow fading, ${{\kappa}_{t,r}}(dB)=0.5\mathrm{dB}$
and ${{\phi}_{t,r}}={{3}^{\circ}}$}).\vspace{-0.5cm} }
\label{fig:coh_dif}
\end{figure}

\begin{figure}[tb]
\centering\includegraphics[width=3.5in]{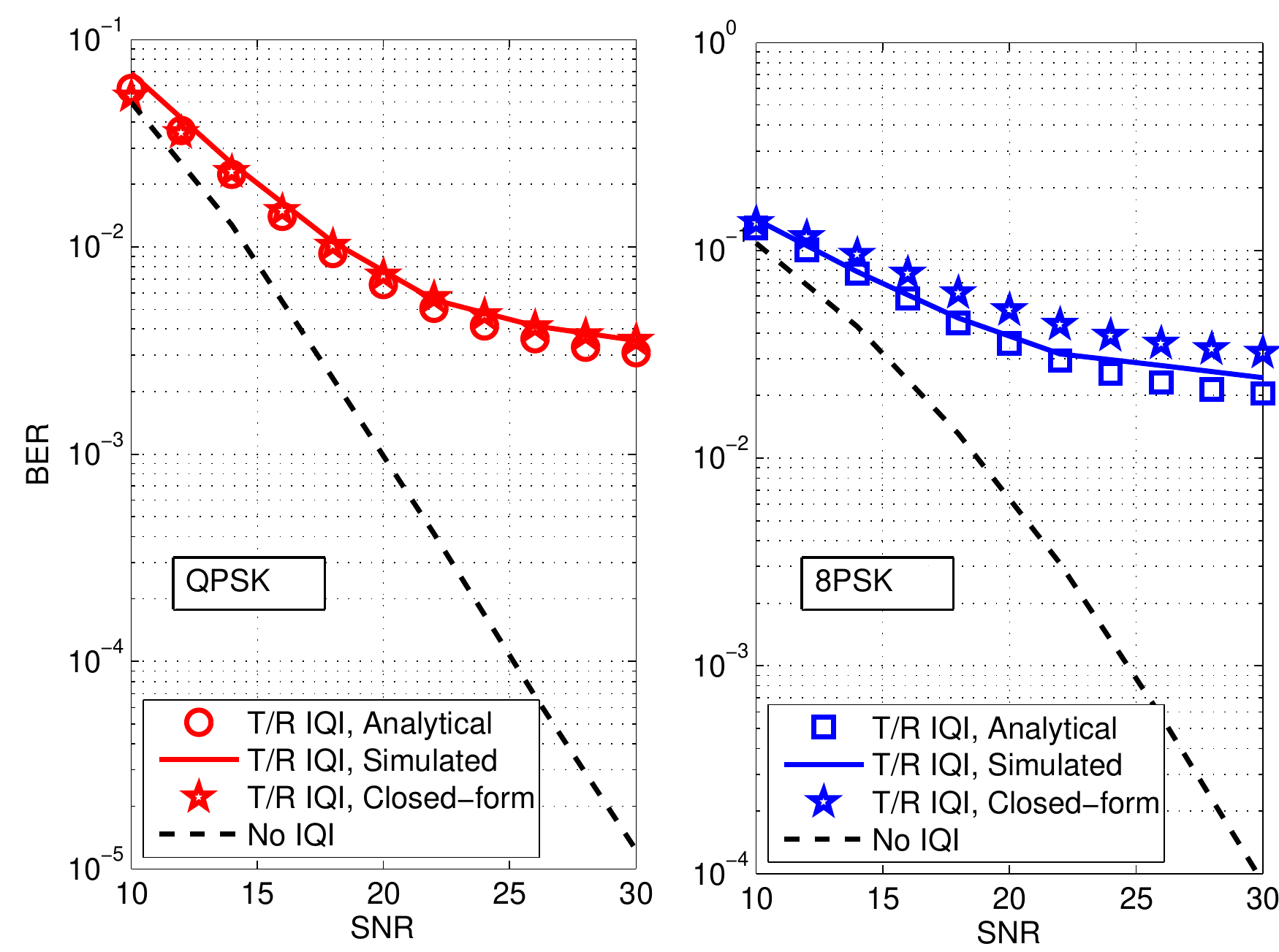}\vspace{-0.5cm}\caption{Comparison of analytical and simulated BER floor of the DSTBC-OFDM
system under TX-/RX-IQI ({\small{}slow fading, ${{\kappa}_{t,r}}(dB)=0.5\mathrm{dB}$
and ${{\phi}_{t,r}}={{3}^{\circ}}$)}\vspace{-0.5cm} }
\label{fig:Performance_TR}
\end{figure}

Fig. \ref{fig:coh_dif} compares the impact of IQI on differential
and coherent detection. The channel information in coherent detection
is assumed to be perfectly known. As expected, when there is no IQI,
the SNR gap between coherent and differential detection is roughly
3dB, and becomes larger in the presence of IQI, especially under RX-IQI.
According to Subsection \ref{cohcompare}, the performance of coherent
detection has an advantage of 3dB in both IRR and SNR. Hence the BER
of coherent detection under RX-IQI can be well predicted by setting
the noise power $\sigma^{2}$ and the IQI interference power $\rho_{t/r}^{(c)}$
to half of their values in the differential system, which is also
shown in the figure. Fig. \ref{fig:Performance_TR} shows the analytical
and closed-form BER of QPSK and 8PSK in the presence of both the TX-IQI
and RX-IQI and compares them with the simulated BER. The analytical
and closed-form BER are calculated by replacing the noise power term
in Eq. \eqref{eq:BER-TX-analytical} and \eqref{eq:BER-TX-closedform}
as $\sigma^{2}\rightarrow(\sigma^{2}+\rho_{r})$.

\begin{figure}[tb]
\centering\includegraphics[width=3.5in]{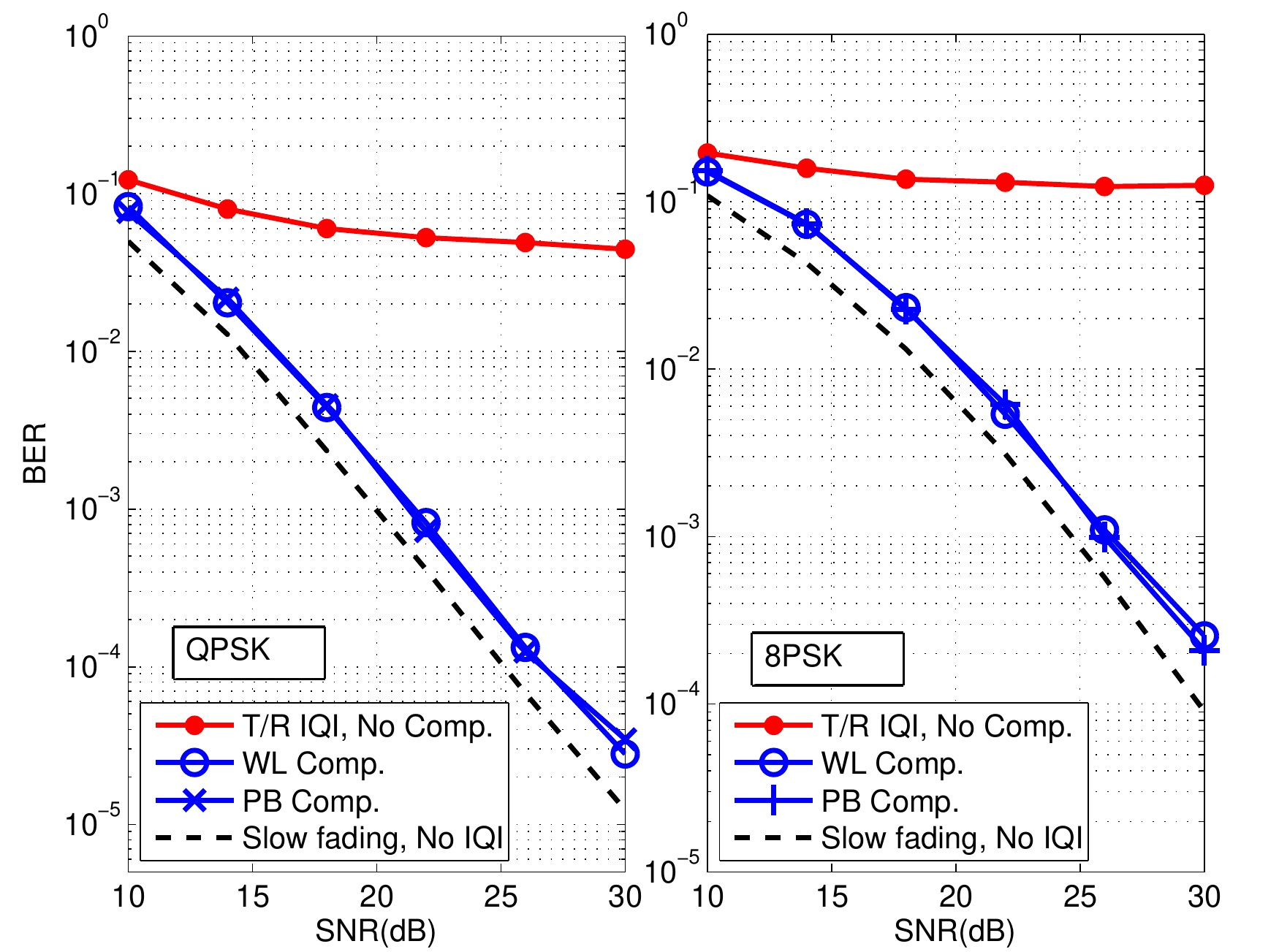}\vspace{-0.5cm}\caption{BER performance of the DSTBC-OFDM system under joint TX-/RX-IQI in
slow fading channels ({\small{}${{\kappa}_{t,r}}(dB)=1\mathrm{dB}$,
${{\phi}_{t,r}}={{5}^{\circ}}$}).\label{tx-slow}\vspace{-0.2cm}}
\end{figure}

\begin{figure}[tb]
\centering \includegraphics[width=3.5in]{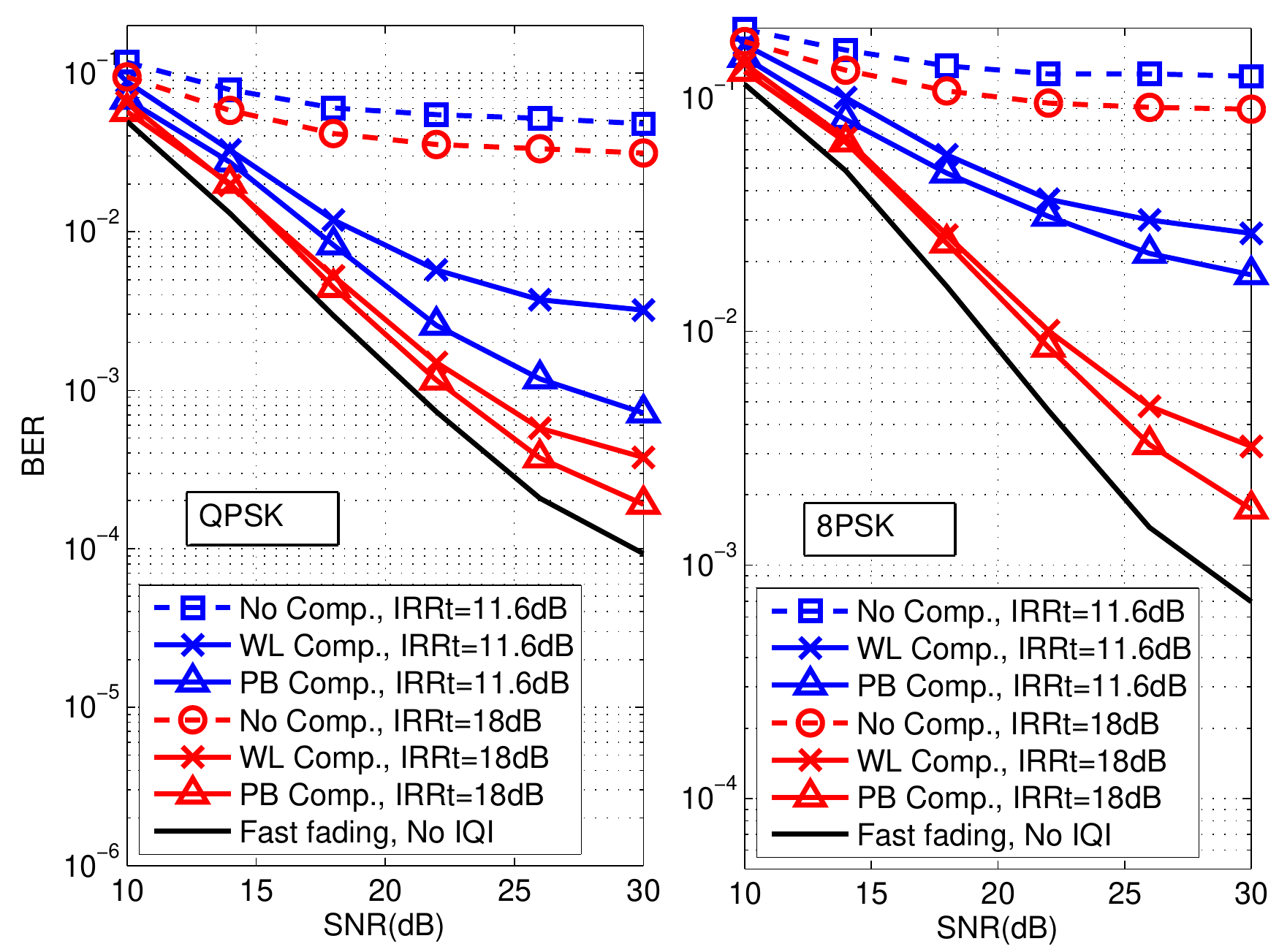} \vspace{-0.5cm}\caption{BER performance of the DSTBC-OFDM system under joint TX-/RX-IQI in
fast fading channels ({\small{}when $IRR_{t}=11.6\text{dB}$,${{\kappa}_{t,r}}(dB)=1\mathrm{dB}$,
${{\phi}_{t,r}}={{5}^{\circ}}$}); when {\small{}$IRR_{t}=18\text{dB}$},{\small{}
${{\kappa}_{t,r}}(dB)=1\mathrm{dB}$, ${{\phi}_{t,r}}={{5}^{\circ}}$}).
\label{TR-fast}\vspace{-0.5cm}}

\end{figure}

Fig. \ref{tx-slow} shows the performance of our proposed blind IQI
compensation algorithms in a slow-fading channel. Both the TX-IQI
and RX-IQI parameters are set to severe IQI with ${{\kappa}_{t/r}}(dB)=1\mathrm{dB}$,
${{\phi}_{t/r}}={{5}^{\circ}}$. Both compensation with the PB estimation
and WL estimation can effectively mitigate the performance degradation
due to IQI. However, we still observe a 1.6dB loss in SNR compared
to the IQI-free case for both QPSK and 8PSK modulation. According
to our analysis in Subsection \ref{subsec:SNR-degradation-TX-comp},
even with the perfect compensation matrix, there will be an inevitable
SNR loss of $10\mathrm{log}\frac{(1+2|\beta_{t}^{2}|/|\alpha_{t}^{2}|)}{(|\alpha_{t}|-|\beta_{t}|^{2}/|\alpha_{t}|)^{2}}=$1.2dB
due to the signal power loss and noise amplification in the compensation
of TX-IQI. The rest of SNR loss is caused by the estimation error
due to noise. Moreover, Fig. \ref{TR-fast} presents the performance
of our proposed blind IQI compensation algorithm in a fast-fading
channel under joint TX-IQI and RX-IQI. The RX-IQI parameters are set
to ${{\kappa}_{r}}(dB)=1\mathrm{dB},{{\phi}_{r}}={{5}^{\circ}}$,
while both the moderate TX-IQI case ($\mathrm{IRR}{}_{t}=18$dB) and
severe TX-IQI case ($\mathrm{IRR}{}_{t}=11.6$dB) are simulated. A
performance degradation is observed in the fast-fading channel even
without IQI since the fast-varying channel does not satisfy the quasi-static
property required by differential STBC. Moreover, it is clear that
the PB estimation outperforms the WL estimation for the fast fading
channel. Significant improvement can be observed in the presence of
severe TX-IQI. On the other hand, regarding the improvement in BER
level after compensation, it is more noticeable for QPSK than that
for 8PSK when compared with the WL estimation. However, the SINR improvement
due to compensation should be basically the same because the compensation
matrices are estimated based on modulated symbols. This is also confirmed
by the fact that the SNR gap between the PB estimation and the WL
estimation is the same for QPSK and 8PSK modulation. 

\begin{figure}[tb]
\noindent \begin{centering}
\centering \includegraphics[width=3.5in]{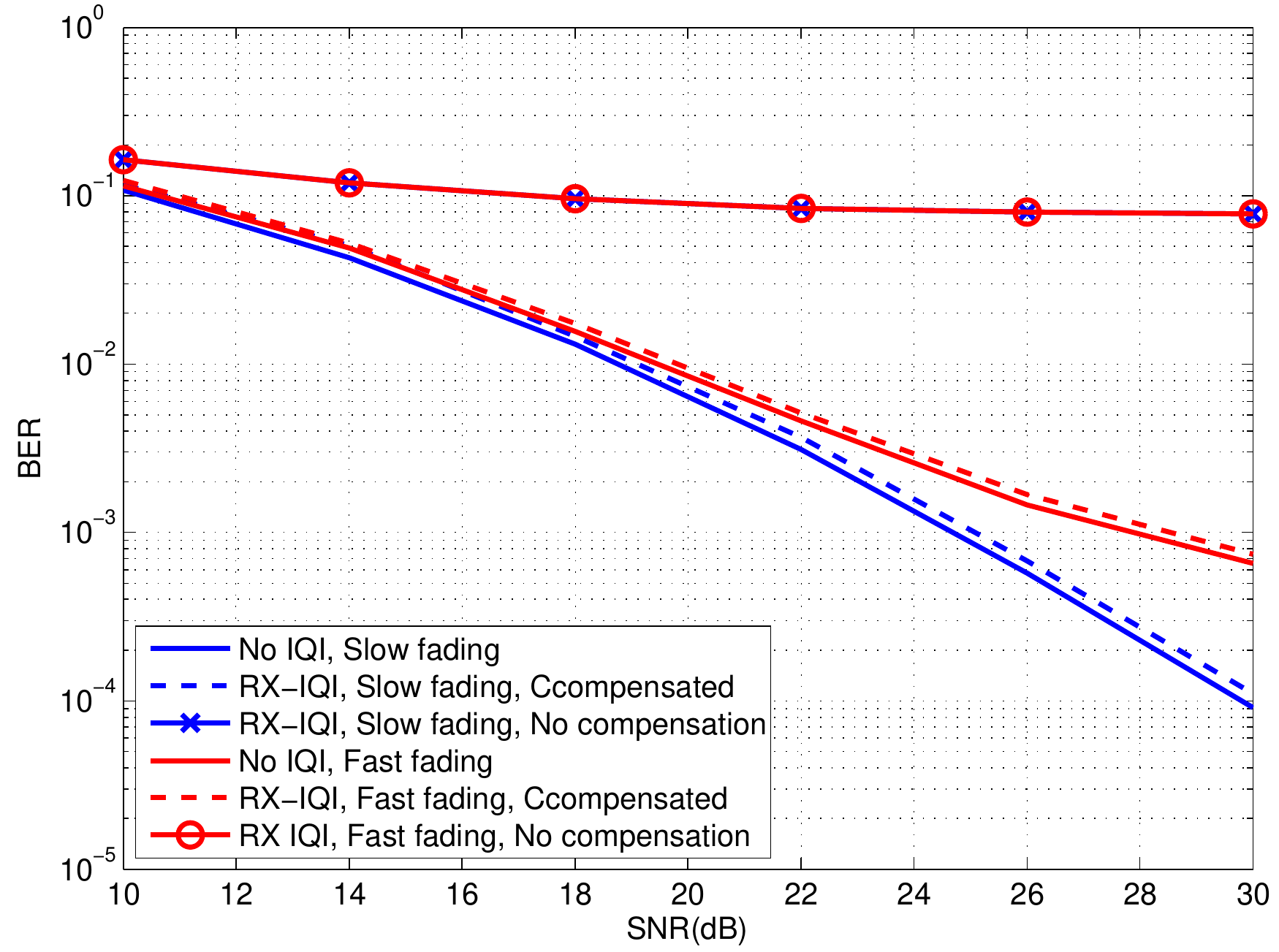} \vspace{-0.5cm}\caption{BER performance of the DSTBC-OFDM system under RX-IQI in slow and
fast fading channels ({\small{}$\kappa_{r}(dB)=1\mathrm{dB}$, $\phi_{r}={{5}^{\circ}}$})\label{Fig rx BEr}\vspace{-0.5cm}}
\par\end{centering}
\end{figure}

The performance of the WL RX-IQI compensation in Subsection \ref{subsec:RX-IQI-only}
is presented in Fig. \ref{Fig rx BEr} for both fast-fading and slow-fading
channels assuming 8PSK modulation. The RX-IQI parameters are set to
${{\kappa}_{r}}(dB)=1\mathrm{dB}$ and ${{\phi}_{r}}={{5}^{\circ}}$.
Fig. \ref{Fig rx BEr} shows that the proposed compensation algorithm
efficiently compensates for RX-IQI. Since the RX-IQI compensation
matrix does not change with the channel, the compensation is effective
in both channel scenarios and the degradation caused by RX-IQI is
almost eliminated.\vspace{0.02cm}

\begin{figure}[tb]
\centering\includegraphics[width=3.5in]{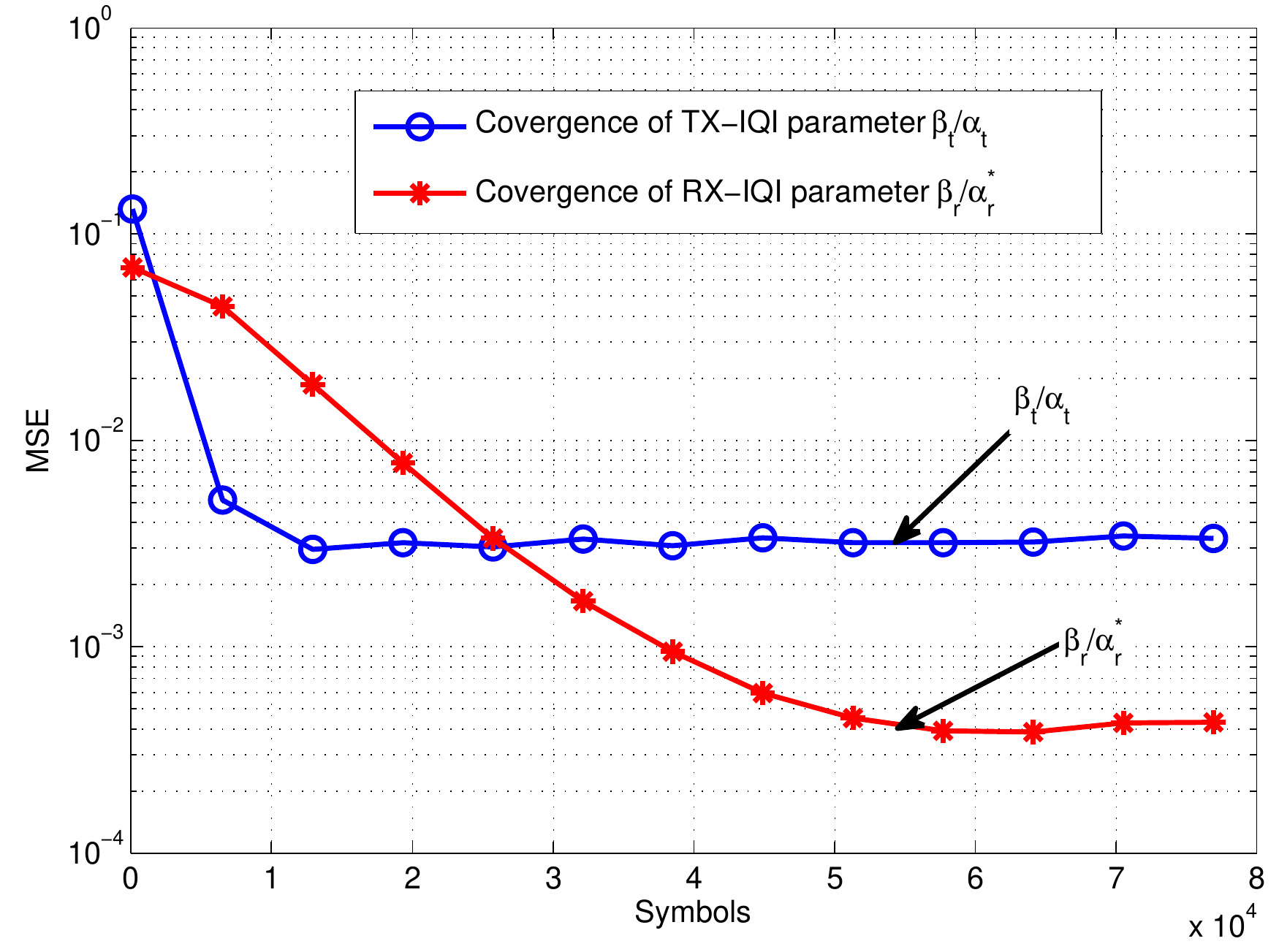} \vspace{-0.5cm}\caption{Convergence rate of TX-/RX-IQI parameter estimation (mobile speed
200km, SNR=3dB)\label{fig: cvg}\vspace{-0.5cm}}

\end{figure}

Next, we examined the convergence speed and mean squared error (MSE)
of the IQI parameter estimation discussed in Subsection \ref{subsec:Estimation-of-TX-IQI}
under severe TX-IQI and RX-IQI in a fast fading channel. The results
are shown in Fig. \ref{fig: cvg}. It can be observed from Fig. \ref{fig: cvg}
that the RX-IQI can be estimated more accurately than the TX-IQI parameters.
This is due to the fact that the estimation of the TX-IQI parameter
is biased by the approximation that the term $\mathbf{B}_{r}^{*}\mathbf{B}_{t}\mathbf{\Lambda}(n)$
in Eq. \eqref{eq:taoc_approx} is ignored.

\section{Conclusions}

\label{sec:conc}\vspace{-0.05cm}In this paper, we analyzed the impact
of TX-IQI and RX-IQI in DSTBC-OFDM. We quantified analytically the
BER increase caused by TX-IQI, the BER floor due to RX-IQI and the
analytical BER curve of M-PSK under TX-IQI and RX-IQI. The accuracy
of these analyses was demonstrated by simulations. In addition, an
adaptive decision-directed joint TX-/RX-IQI compensation algorithm
for DSTBC-OFDM was proposed and demonstrated to effectively mitigate
the performance degradation caused by IQI. We also proposed an enhancement
for the high-mobility case by exploiting constrained relationship
among a pair of image OFDM subcarriers.

\appendices{\vspace{-0.1cm}}

\section{BER floor condition under TX-IQI\label{subsec:Apa}}

Since ${{\mathbf{S}}_{k}}{{\mathbf{S}}_{k}}^{H}={{\mathbf{\bar{S}}}_{k}}{{\bar{\mathbf{S}}}_{k}}^{H}=\mathbf{I}$
and ${\mathbf{S}}_{k}$ and ${\mathbf{\bar{S}}}_{k}$ have Alamouti
structure, the entries in ${\mathbf{S}}_{k}$ and ${\mathbf{\bar{S}}}_{k}$
satisfy the power constraint of $\left|[{\mathbf{S}}_{k}]_{m,n}\right|^{2}\leq1,\forall m,n\in[1,2].$
Therefore, according to Eq. \eqref{eq:detectionmetric-t}, the entries
of the interference matrix $\Theta_{TX}=|{{\alpha}_{t}}{{\beta}_{t}}|{{\mathbf{S}}_{k}}^{H}\mathbf{{\mathbf{\bar{S}}}}_{k+1}+|{{\alpha}_{t}}{{\beta}_{t}}|{{\mathbf{\bar{S}}}_{k}}^{H}\mathbf{{\mathbf{S}}}_{k+1}$
should satisfy the power constraint: $\left|[\Theta_{TX}]_{m,n}\right|^{2}\leq2|{{\alpha}_{t}}{{\beta}_{t}}|,\forall m,n\in[1,2]$,
which means that the largest error vector magnitude (EVM) introduced
by TX-IQI is limited to $2|{{\alpha}_{t}}{{\beta}_{t}}|$. On the
other hand, given an M-PSK constellation, the minimum EVM needed to
cause a bit error, defined as $A_{min}$, is determined by the distance
from a given symbol to the detection boundary of its neighbor in the
signal constellation. For 8-PSK signal with normalized power, $A_{min}=\sin\frac{\pi}{8}$.
Thus, when the largest EVM introduced by TX-IQI is smaller than $A_{min}$,
there will be no error as $\mathrm{SNR}\rightarrow\infty$. In DSTBC-OFDM,
the transmitted symbol is normalized by a factor of $\sqrt{2}$, thus
the minimum EVM needed to cause a bit error in DSTBC-OFDM is $\frac{A_{min}}{\sqrt{2}}$.
Consequently, TX-IQI will not introduce error as $SNR\rightarrow\infty$
unless the maximum error power is larger than $\frac{A_{min}}{\sqrt{2}}$,
$i.e.$ $|{{\alpha}_{t}}{{\beta}_{t}}|<\frac{\sqrt{2}}{4}A_{min}$.

\section{Derivation of equivalent amplitude loss under TX-IQI\label{subsec:Apb}}

Assume, without loss of generality, that the symbol $u=1/\sqrt{2}$
is transmitted. The symbol error probability could be approximated
by the pairwise error probability{\small{}
\begin{equation}
P_{e,u}\approx P_{e}(u\rightarrow\frac{2\pi}{M})+P_{e}(u\rightarrow-\frac{2\pi}{M})
\end{equation}
}where $P_{e}(u\rightarrow\frac{2\pi}{M}$) and $P_{e}(u\rightarrow-\frac{2\pi}{M})$
are the probabilities that the transmitted symbol $u$ is detected
as $\frac{\pi}{M}$ and $-\frac{\pi}{M}$ at the receiver, respectively.
As shown in Fig. \ref{fig: signal TX}, an error will occur when the
received symbol falls outside the detection boundaries $\overline{OA}$
and $\overline{OB}$ due to the noise effect. In the high SNR case,
the probability of error $P_{e}(u\rightarrow\frac{2\pi}{M}$) can
be approximated by the probability that the received symbol falls
to the other side of line $\overline{OA}$. Likewise, the symbol error
probability $P_{e}(u\rightarrow-\frac{2\pi}{M}$) can also be approximated
by the probability that the received symbol falls to the other side
of line $\overline{OB}$. Under these approximations, the pairwise
error probability $P_{e}(u\rightarrow\frac{2\pi}{M}$) and $P_{e}(u\rightarrow-\frac{2\pi}{M})$
is totally determined by the distance from the symbol $u$ to the
detection boundaries $\overline{OA}$ and $\overline{OB}$, which
are denoted as $D(u,\overline{OA})$ and $D(u,\overline{OB})$, respectively.
Hence, when TX-IQI introduces an error vector $\varepsilon$ and a
gain $|\alpha_{t}|^{2}$ to the transmitted signal, $i.e$. $u'=|\alpha_{t}|^{2}u+\varepsilon$,
the distances between the transmitted symbol and the decision boundaries
are changed, denoted as $D(u',\overline{OA})$ and $D(u',\overline{OB})$,
so the pairwise error probability becomes $P_{e,u'}\approx P_{e}(u'\rightarrow\frac{2\pi}{M})+P_{e}(u'\rightarrow-\frac{2\pi}{M})$.
Define a pair of equivalent signal amplitude losses $\epsilon^{+}\in\mathbf{R}$
and $\epsilon^{-}\in\mathbf{R}$ that satisfy $P_{e}(u'\rightarrow\frac{2\pi}{M})=P_{e}\left((u-\epsilon^{+})\rightarrow\frac{2\pi}{M}\right)$
and $P_{e}(u'\rightarrow-\frac{2\pi}{M})=P_{e}\left((u-\epsilon^{-})\rightarrow-\frac{2\pi}{M}\right)$,
respectively. According to our previous analysis, the distances from
$(u-\epsilon^{+})$ and $(u-\epsilon^{-})$ to the corresponding detection
boundary should be equal to that of $u'$, $i.e.$ $D(u-\epsilon^{+},\overline{OA})=D(u',\overline{OA})$
and $D(u-\epsilon^{-},\overline{OB})=D(u',\overline{OB})$. Thus,
after some geometrical calculations, $\epsilon^{+}$ and $\epsilon^{-}$
are given by{\small{}
\begin{eqnarray}
\epsilon^{+} & = & |a_{\varepsilon}|\sin(\varphi_{\varepsilon})\cot(\pi/M)-|a_{\varepsilon}|\cos(\varphi_{\varepsilon})\nonumber \\
\epsilon^{-} & = & |a_{\varepsilon}|\sin(\varphi_{\varepsilon})\cot(\pi/M)+|a_{\varepsilon}|\cos(\varphi_{\varepsilon})
\end{eqnarray}
}where $\varphi_{\varepsilon}$ and $|a_{\varepsilon}|$ are the angle
and amplitude of the TX-IQI error vector $\varepsilon$. According
to our analysis in Section \ref{sec:ber}, $\varepsilon$ is a zero-mean
complex Gaussian random variable with variance $|\alpha_{t}\beta_{t}|^{2}$,
and $|a_{\varepsilon}|\cos(\varphi_{\varepsilon})$ and $|a_{\varepsilon}|\sin(\varphi_{\varepsilon})$
are the real and imaginary parts of $\varepsilon$, respectively.
Hence, they are both zero-mean Gaussian variables with variance $|\alpha_{t}\beta_{t}|^{2}/2$.
Consequently, both $\epsilon^{+}$and $\epsilon^{-}$ are real zero-mean
Gaussian variable with variance $\left(1+\cot(\pi/M){}^{2}\right)|\alpha_{t}\beta_{t}|^{2}$.
Thus, the signal power due to TX-IQI can be expressed as follows{\small{}
\begin{eqnarray}
E_{s,TX} & = & (|\alpha_{t}|^{2}/\sqrt{2}-\epsilon)^{2}\nonumber \\
 & = & (|\alpha_{t}|^{2}/\sqrt{2}+\sqrt{\left(1+\cot(\pi/M){}^{2}\right)|\alpha_{t}\beta_{t}|^{2}/2}n')^{2}\nonumber \\
 & = & (|\alpha_{t}|^{2}+\sqrt{\left(1+\cot(\pi/M){}^{2}\right)|\alpha_{t}\beta_{t}|^{2}}n')^{2}/2
\end{eqnarray}
}where $\epsilon=\sqrt{\left(1+\cot(\pi/M){}^{2}\right)|\alpha_{t}\beta_{t}|^{2}/2}n'$
is the random amplitude loss and $n$ is a real zero-mean Gaussian
random variable with variance 1.

\bibliographystyle{IEEEtran}

\begin{IEEEbiography}[{\includegraphics[width=1in,height=1.25in]{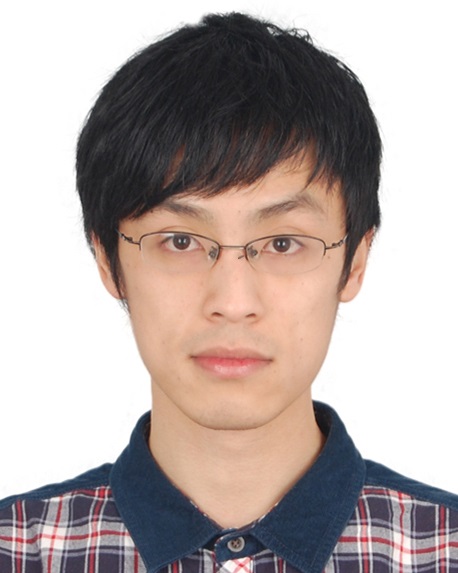}}]{Lei Chen}
 received the B.Eng. degree in communication engineering from UESTC,
China, in 2010. He is currently a Ph.D. candidate in the National Key Laboratory of Science and Technology on Communications at UESTC. He joined Dr. Naofal Al-Dhahir's group in University of Texas at Dallas as a visiting Ph.D. student in 2015-2016. His research
interests are focused on millimeter-wave communications, especially on the baseband compensation of radio frequency impairments. 
\end{IEEEbiography}
\vspace{-0.5cm}
\begin{IEEEbiography}[{\includegraphics[width=1in,height=1.25in]{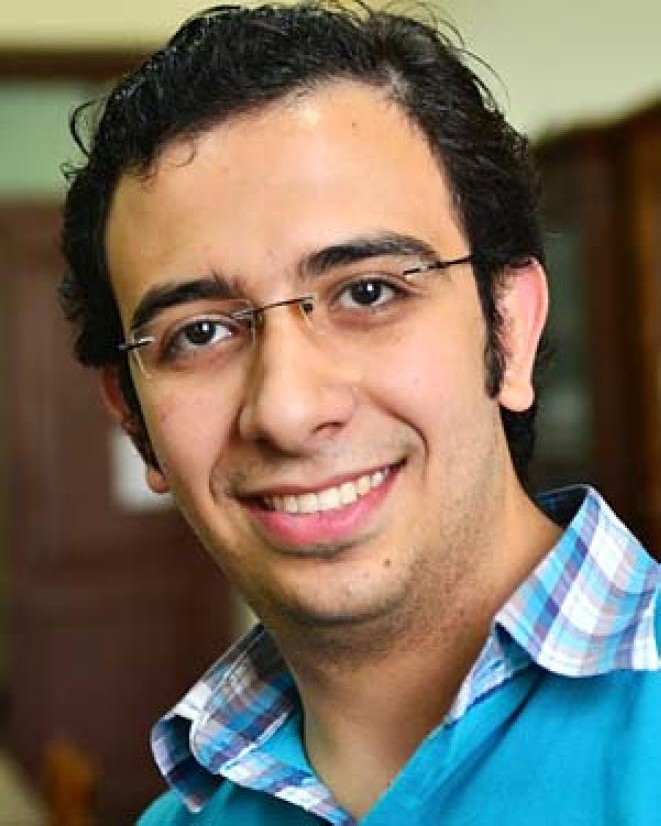}}]{Ahemd G. Helmy} received the B.Sc. and M.Sc. degrees with honors in electronics and communications engineering from Cairo University, Egypt, in 2008 and 2011, respectively. Currently, he is pursuing his Ph.D degree in Electrical Engineering in wireless communication in the University of Texas at Dallas. His research interests include Digital Signal Processing, Interference Mitigation Techniques for Multi-user MIMO systems, and their applications to various wireless standards, like WLAN and LTE. Calling the industrial background, he had many experiences from working for Apple Inc., Intel Corp., Xtendwave Semiconductors, Wasiela Semiconductors, and different R\&D research projects collaborated with many industrial leading labs for IMEC, Vodafone, and Qtel. During that time, he focused on how to homogeneously merge his theoretical background and practical experience together in a tangible output emphasized through ling several patents.
\end{IEEEbiography}
\vspace{-0.5cm}
\begin{IEEEbiography}[{\includegraphics[width=1in,height=1.25in]{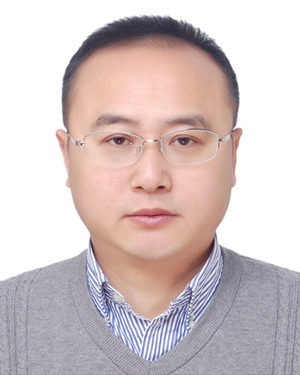}}]{Guangrong Yue}
 received the Ph.D. degrees in communication and information system
from UESTC, Chengdu, China, in 2006. He was a Post-Doctoral Fellow
at the Department of EECS at University of California, Berkeley from 2007 to 2008. He is now a professor of the National
Key Laboratory of Science and Technology on Communications, UESTC.  As one of the  key researchers, he has participated in the Project of Millimeter Wave and Tera-hertz Key Technology and High-speed Baseband Signal Processing supported by the National High-tech R\&D Program of China (863 Program). His major research interests include mobile communications and millimeter-wave communications.
\end{IEEEbiography}
\vspace{-0.5cm}
\begin{IEEEbiography}[{\includegraphics[width=1in,height=1.25in]{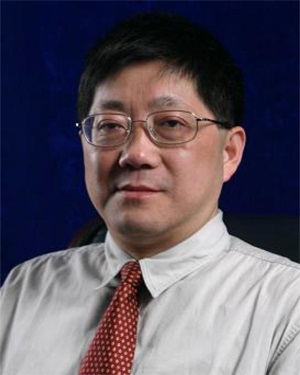}}]{Shaoqian Li}
 (F'16) received the B.S.E. degree in communication technology from
Northwest Institute of Telecommunication (Xidian University), China,
in 1982, and the M.S.E. degree in communication system from UESTC,
China, in 1984. He is a Professor, Ph.D. Supervisor, and the Director
of the National
Key Laboratory of Science and Technology on Communications, UESTC, and a member of the National
High Technology R\&D Program (863 Program) Communications Group. His research interests include wireless communication theory and anti-interference
technology.
\end{IEEEbiography}
\vspace{-0.5cm}
\begin{IEEEbiography}[{\includegraphics[width=1in,height=1.25in]{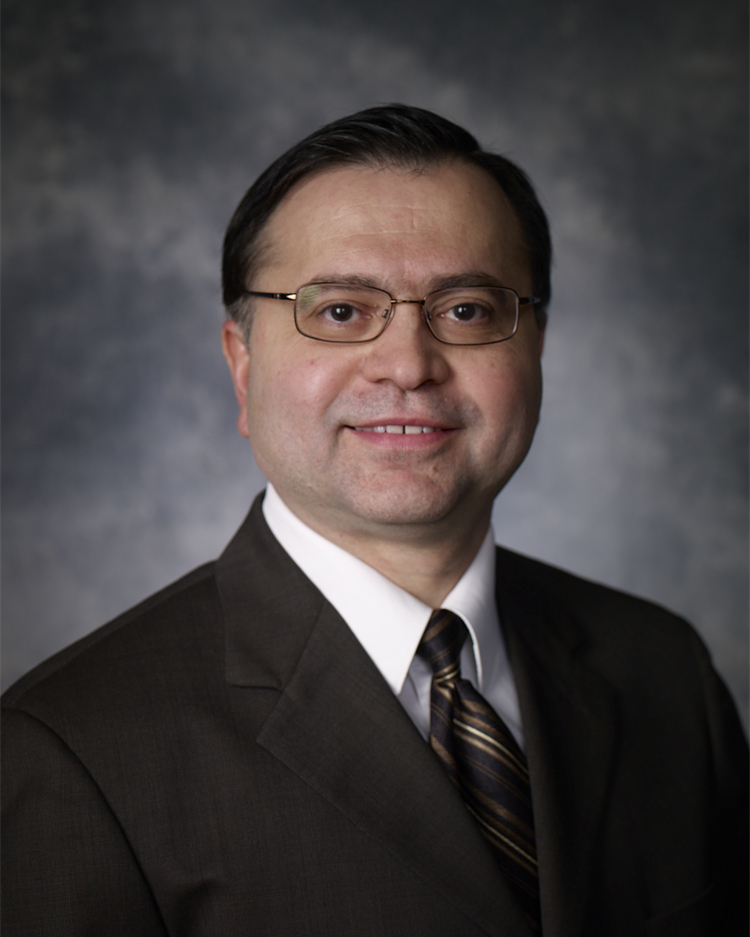}}]{Naofal Al-Dhahir} (S'89-M'90-SM'98-F'08) received the Ph.D. degree in electrical engineering from Stanford University. He is the Erik Jonsson Distinguished Professor with The University of Texas at Dallas. From 1994 to 2003, he was a Principal Member of the Technical Staff with GE Research and AT\&T Shannon Laboratory. He is co-inventor of 41 issued U.S. patents, co-author of over 325 papers with over 8000 citations, and co-recipient of four IEEE best paper awards. He is the Editor-in-Chief of the IEEE TRANSACTIONS ON COMMUNICATIONS.
\end{IEEEbiography}
\end{document}